\documentclass[twocolumn]{aastex63}

\accepted{for publication in ApJ, May 2021}
\shorttitle{Pre-Main Sequence X-ray Super-Flares}
\shortauthors{Getman \& Feigelson}

\graphicspath{{./}}

\begin{document}

\title{X-ray Super-Flares From Pre-Main Sequence Stars: Flare Energetics And Frequency}

\correspondingauthor{Konstantin Getman}
\email{kug1@psu.edu}

\author[0000-0002-6137-8280]{Konstantin V. Getman}
\affiliation{Department of Astronomy \& Astrophysics \\
Pennsylvania State University \\ 
525 Davey Laboratory \\
University Park, PA 16802, USA}

\author{Eric D. Feigelson}
\affiliation{Department of Astronomy \& Astrophysics \\
Pennsylvania State University \\ 
525 Davey Laboratory \\
University Park, PA 16802, USA}
\affiliation{Center for Exoplanetary and Habitable Worlds}

\begin{abstract}
Solar-type stars exhibit their highest levels of magnetic activity during their early convective pre-main sequence (PMS) phase of evolution. The most powerful PMS flares, super-flares and mega-flares, have peak X-ray luminosities of $\log(L_X)=30.5-34.0$~erg~s$^{-1}$ and total energies $\log(E_X)=34-38$~erg. Among $>24,000$ X-ray selected young ($t\la5$~Myr) members of 40 nearby star-forming regions from our earlier $Chandra$ MYStIX and SFiNCs surveys, we identify and analyze a well-defined sample of 1,086 X-ray super-flares and mega-flares, the largest sample ever studied. Most are considerably more powerful than optical/X-ray super-flares detected on main sequence stars. This study presents energy estimates of these X-ray flares and the properties of their host stars. These events are produced by young stars of all masses over evolutionary stages ranging from protostars to diskless stars, with the occurrence rate positively correlated with stellar mass. Flare properties are indistinguishable for disk-bearing and diskless stars indicating star-disk magnetic fields are not involved. A slope $\alpha\simeq2$ in the flare energy distributions $dN/dE_X \propto E_X^{-\alpha}$ is consistent with those of optical/X-ray flaring from older stars and the Sun. Mega-flares ($\log(E_X) > 36.2$~erg) from solar-mass stars have occurrence rate of $1.7_{-0.6}^{+1.0}$ flares/star/year and contribute at least $10-20$\% to the total PMS X-ray energetics. These explosive events may have important astrophysical effects on protoplanetary disk photoevaporation, ionization of disk gas, production of spallogenic radionuclides in disk solids, and hydrodynamic escape of young planetary atmospheres. Our following paper details plasma and magnetic loop modeling of the $>50$ brightest X-ray mega-flares.
\end{abstract}

\section{Introduction} \label{sec:intro}

\subsection{Pre-Main Sequence Super-flares and Mega-flares}

X-ray imaging studies of nearby star forming regions, such as the Taurus clouds and Orion Nebula, typically show that highly variable X-ray emission is a ubiquitous characteristic of pre-main sequence (PMS) stars \citep{Feigelson1981,Montmerle1983,Getman05,Gudel07}. The emission arises from magnetic reconnection events similar to, but much more powerful and frequent than, flares on the contemporary Sun.  Reviews relating to PMS X-ray emission are provided by \citet{Feigelson1999}, \citet{Gudel2004}, \citet{Feigelson2007}, \citet{Gregory2010}, \citet{Stelzer2017}, \citet{Feigelson2018}, \citet{Sciortino2019} and \citet{Argiroffi2019}.

Though PMS X-ray flares were surprising at first, the existence of strong magnetic dynamos in the interiors of fully convective, rapidly rotating stars, followed by eruption of field lines and violent magnetic reconnection above the stellar surface, is reasonable.  The X-ray emission seems to be independent of the presence or absence of  protoplanetary disks despite astrophysical calculations that star-disk magnetic field lines may be involved in X-ray emitting flares \citep{Hayashi1996,Shu1997,Aarnio2010,Lopez-Santiago2016,Colombo2019}. A factor of two reduced X-ray activity level in accreting versus non-accreting systems  \citep{Flaccomio2003,Preibisch05,Telleschi07} could have several possible causes: cooling of active regions by accreting material,  attenuation of X-rays by accreting columns and/or inner disks, coronal stripping by disks, and distortion of magnetic topologies by disks/accretion \citep{Jardine2006, Gregory2007, Flaccomio2003, Getman08b, Flaccomio2012}.  Accretion shocks contribute a small fraction to the total X-ray emission from T-Tauri stars in the form of soft X-ray excess emission \citep{Telleschi2007b}. 

Following previous researchers \citep{Favata2005,Getman08a,McCleary2011}, we focus attention here on the most luminous PMS X-ray flares with peak X-ray luminosities exceeding $L_{X,pk} = 10^{30.5}$ erg s$^{-1}$ and/or  total (time-integrated) energies exceeding $E_X = 10^{34}$ erg.  In contrast, no solar flare has been observed with total X-ray energy exceeding $\sim 10^{30}$ erg, four orders of magnitude below our threshold \citep{Schrijver2012}. We call events with $10^{34} < E_X < 10^{36.2}$~erg 'super-flares' and events with $E_X > 10^{36.2}$~erg 'mega-flares'\footnote{This boundary represents our completeness limit: All mega-flares in the observed stars have been confidently detected, while only an incomplete subset of super-flares are found (\S~\ref{sec:energy_distributions}).}, recognizing that the super-flare designation is also used for less powerful optical flares seen with the Kepler satellite in older stars. In both solar and PMS flares, X-rays constitute only a minor fraction of the total radiated energy \citep{Flaccomio2018}, and the radiated energy may be dominated by the energy in ejected magnetic fields and energetic particles.

In the present effort, we examine the largest sample of X-ray PMS super- and mega-flares ever collected to seek ensemble characteristics and relationships with other properties.  The sample is drawn from observations of $>24,000$ PMS stars detected in 40 MYStIX \citep[Massive Young Star-forming complex study in Infrared and X-rays;][]{Feigelson13} and SFiNCs \citep[Star Formation in Nearby Clouds;][]{Getman17} star forming regions with NASA's {\it Chandra} X-ray Observatory (\S\ref{sec:datasets}). 

\subsection{Astrophysical Implications} 

Our studies are aimed at partially addressing some of the important questions concerning flare related physics and phenomena:     

\begin{enumerate}

\item Previous studies using much smaller samples of X-ray flares from PMS stars \citep{Wolk05, Stelzer07, Caramazza07, Colombo07} report similar powerlaw slopes $\alpha \sim 2$ of flare energy distributions $dN/dE_{flare} \propto E_{flare}^{-\alpha}$ consistent with those of older stars and the Sun. Such findings may have implications for understanding  the relative importance of powerful flares and nano- or micro-flares for heating  the  solar  and  stellar  coronas \citep[][and references therein]{Nhalil2020}.  

\item By taking advantage of the much increased super-flare sample combined with a homogeneous set of derived flare-host properties (such as stellar mass), we evaluate flare occurrence rate as a function of flare energy and stellar mass, as well as contribution of  powerful flares to the total X-ray fluence of PMS stars. Such unique estimates will provide better understanding the effects of PMS X-ray emission on their molecular environs including their natal molecular cloud, the infalling envelope of protostars, the protoplanetary disk around T-Tauri stars, and the protoplanets revealed after the disk has dissipated. 

\item Super- and mega-flares are important for both their high fluency ionizing radiation and the production of hard X-rays that can penetrate deep into molecular environments \citep{Glassgold2000}.  Flare X-rays potentially can produce layers of ionization in otherwise neutral material, induce non-equilibrium ion-molecular chemistry, and sputter grain surfaces. Even low levels of ionization can couple molecular material to magnetic fields resulting (in some circumstances) in turbulent motions and (in other circumstances) in bulk outflows. There is some empirical evidence that PMS flares heat disks and diminish accretion due to photoevaporation of disks \citep{Drake2009,Flaccomio2018,Flaischlen2021}. They may be accompanied by energetic particles that could produce spallogenic radionuclides and by coronal mass ejection shocks that could melt ices or solids. 

Flare radiation play an important role in photoevaporative flows and dispersal of protoplanetary disks \citep[reviewed by][]{Williams2011, Alexander2014, Ercolano2017} and young planetary atmospheres \citep{Lammer2003, Ribas2005, Gudel2007, Gronoff20}.  There is particular concern that the effects of violent magnetic activity in young stars $-$ extreme ultraviolet emission and coronal mass ejections as well as X-ray emission $-$ can erode atmospheres of planets that otherwise might be habitable \citep{Lammer07, Gronoff20, Atri20}.  The effects of energetic particles and coronal mass ejections that may be associated with super-flares is still uncertain \citep{Drake2016,Atri2020b}.

\item The geometry of PMS flare plasma seems remarkable. Models of X-ray evolution of PMS super-flares are usually consistent with enormous loop structures often larger than the star itself \citep{Favata2005,Getman08b,Reale2018}.  Magnetospheric calculations for classical T-Tauri stars indicate that closed magnetic loops on this scale can co-exist with open magnetic field lines accreting from a protoplanetary disk \citep[][and references therein]{Johnstone2014}.  However it is difficult to exclude other magnetic geometries such as more complicated sequentially triggered arcades \citep{Getman2011} or eclipsed loop geometries \citep{Johnstone2012}. 

\item The astrophysics of flare plasma needs to be investigated.  Are the heating and cooling processes similar to those of solar flares even when the emission measure is vastly greater? To address some of these issues, in our following paper (Getman, Feigelson, \& Garmire, 2021, ApJ, submitted) we perform detailed modeling of the brightest MYStIX/SFiNCs X-ray super-flares from disky and diskless PMS stars of various masses, involving evaluation of plasma temperature and emission measure temporal profiles, flare cooling time-scales, coronal loop length and thickness, and further comparison of these properties with those of numerous X-ray flares from young stars in Orion Nebula \citep{Getman08a} and flares from older stars \citep{Gudel2004} and the Sun \citep{Aschwanden2008}.  

\item The explosive and thermal processes within the magnetic structures producing super- and mega-flare plasma needs to be investigated.  Are the heating and cooling processes similar to those of solar flares even when the emission measure is vastly greater?  Some flares have temporal profiles that differ from the classic ``fast rise exponential decay'' behavior common among solar and stellar flares \citep{Getman08a}.  Do these reflect different loop morphologies  or different heating or cooling mechanisms? 

\end{enumerate}

\subsection{Outline of the Paper} 

After describing the {\it Chandra} datasets (\S \ref{sec:datasets}), our procedures for selecting, classifying, and characterizing X-ray super-flares are described in section \S \ref{sec:methods} and the Appendices.  Section \S \ref{sec:results} presents the distributions of flare luminosity and energy.
Section \S \ref{sec:allmystixsfincs_vs_super-flares} provides dependencies of super- and mega-flare properties on stellar mass and presence/absence of circumstellar disk. Super- and mega-flare occurrence rates as functions of flare energy and stellar mass, as well as their contribution to the total X-ray fluence of PMS stars are evaluated in \S \ref{sec:flare_frequency}. Comparison of X-ray and optical super-flares is given in \S \ref{sec_optical}. Effects of super- and mega-flares on the environment of young stars are discussed in \S \ref{sub_sec_d3}. Concluding remarks are presented in \S \ref{sec_concluding_remarks}.
  
\begin{deluxetable*}{crrrrrc|crrrrr}
%\tablenum{1}
\tabletypesize{\footnotesize}
\tablecaption{MYStIX and SFiNCs Regions \label{tab:mystix_sfincs_regions}}
\tablewidth{0pt}
\tablehead{
\colhead{Region} & \colhead{R.A.} &
\colhead{Dec.} & \colhead{Dis.}& \colhead{$N_{X,obs}$} & \colhead{$N_{tot}$} &&
\colhead{Region} & \colhead{R.A.} &
\colhead{Dec.} & \colhead{Dis.}& \colhead{$N_{X,obs}$} & \colhead{$N_{tot}$} \\
\colhead{} & \colhead{(deg)} & \colhead{(deg)} & \colhead{(pc)} & \colhead{} & \colhead{} &&
\colhead{} & \colhead{(deg)} & \colhead{(deg)} & \colhead{(pc)} & \colhead{} & \colhead{} \\
\colhead{(1)} & \colhead{(2)} & \colhead{(3)} & \colhead{(4)} & \colhead{(5)} & \colhead{(6)} &&
\colhead{(1)} & \colhead{(2)} & \colhead{(3)} & \colhead{(4)} & \colhead{(5)} & \colhead{(6)}
}
\startdata
Be 59 & 0.6 & 67.4 & 1100~~~~ & 464~~~ & 1703~~ && NGC 2264 & 100.3 & 9.6 & 738~~~~ & 898~~~ & 1837~~\\ 
BRC 2 & 1.0 & 68.5 & 1100~~~~ & 42~~~ & 142~~ && NGC 2362 & 109.7 & -25.0 & 1332~~~~ & 467~~~ & 512~~\\
Carina Neb. & 161.2 & -59.7 & 2620~~~~ & 6751~~~ & 37899~~ && NGC 3576 & 168.0 & -61.2 & 2800~~~~ & 1131~~~ & 12235~~\\
Cep A & 344.1 & 62.0 & 868~~~~ & 194~~~ & 534~~ && NGC 6334 & 260.1 & -35.9 & 1770~~~~ & 1385~~~ & 12661~~\\
Cep B & 343.9 & 62.6 & 868~~~~ & 1032~~~ & 1773~~ && NGC 6357 & 261.4 & -34.3 & 1770~~~~ & 1952~~~ & 12581~~\\
Cep C & 346.5 & 62.5 & 868~~~~ & 95~~~ & 247~~ && NGC 7160 & 328.5 & 62.6 & 961~~~~ & 134~~~ & 157~~\\
DR 21 & 309.8 & 42.3 & 1500~~~~ & 594~~~ & 3907~~ && OMC 2-3 & 83.9 & -5.1 & 390~~~~ & 287~~~ & 402~~\\
Eagle Neb. & 274.7 & -13.8 & 1740~~~~ & 2065~~~ & 7762~~ && ONC Flank N & 83.8 & -4.8 & 392~~~~ & 198~~~ & 230~~\\
Flame Neb. & 85.4 & -1.9 & 414~~~~ & 422~~~ & 596~~ && ONC Flank S & 83.8 & -5.7 & 395~~~~ & 223~~~ & 261~~\\
GGD 12-15 & 92.7 & -6.2 & 830~~~~ & 141~~~ & 300~~ && Orion Neb. & 83.8 & -5.4 & 405~~~~ & 1414~~~ & 1700~~\\
IC 348 & 56.1 & 32.1 & 324~~~~ & 307~~~ & 284~~ && RCW 120 & 258.1 & -38.5 & 1680~~~~ & 262~~~ & 1570~~\\
IC 5146 & 328.4 & 47.3 & 783~~~~ & 161~~~ & 272~~ && RCW 36 & 134.9 & -43.8 & 930~~~~ & 337~~~ & 1228~~\\
IRAS 20050 & 301.8 & 27.5 & 700~~~~ & 213~~~ & 308~~ && RCW 38 & 134.8 & -47.5 & 1700~~~~ & 813~~~ & 5750~~\\
LDN 1251B & 339.7 & 75.2 & 300~~~~ & 39~~~ & 39~~ && Rosette Neb. & 98.1 & 4.9 & 1560~~~~ & 1337~~~ & 5029~~\\
Lagoon Neb. & 271.0 & -24.4 & 1336~~~~ & 1828~~~ & 5058~~ && Serpens Main & 277.5 & 1.2 & 440~~~~ & 92~~~ & 149~~\\
Lkh$\alpha$ 101 & 67.5 & 35.3 & 564~~~~ & 197~~~ & 311~~ && Serpens South & 277.5 & -2.0 & 460~~~~ & 78~~~ & 154~~\\
M 17 & 275.1 & -16.2 & 1680~~~~ & 2296~~~ & 13412~~ && Sh 2-106 & 306.9 & 37.4 & 1400~~~~ & 160~~~ & 764~~\\
Mon R2 & 91.9 & -6.4 & 948~~~~ & 410~~~ & 882~~ && Trifid Neb. & 270.6 & -23.0 & 1264~~~~ & 418~~~ & 1279~~\\
NGC 1333 & 52.3 & 31.3 & 296~~~~ & 116~~~ & 111~~ && W 3 & 36.5 & 62.1 & 2040~~~~ & 1571~~~ & 10274~~\\
NGC 1893 & 80.7 & 33.4 & 3790~~~~ & 1110~~~ & 5977~~ && W 4 & 38.2 & 61.5 & 2091~~~~ & 411~~~ & 1614~~\\
NGC 2068 & 86.7 & 0.1 & 414~~~~ & 231~~~ & 335~~ && W 40 & 277.9 & -2.1 & 500~~~~ & 195~~~ & 425~~\\
\enddata
\tablecomments{Column 1: Star forming region. Columns 2-3: Region's approximate position for epoch J2000.0. Column 4: Distance from the Sun. For several regions (DR~21, Flame Nebula, GGD~12-15, IRAC~20050,  LDN~1251B, NGC~2068, NGC~3576, Sh~2-106, W~3, and W~40) we assume the original MYStIX/SFiNCs distances from \citet{Feigelson13,Getman17}. For other regions $Gaia$-based distances are adopted: for RCW~36 from \citet{Fissel2019}; for OMC~2-3, ONC Flanking fields, and Orion Nebula from \citet{Getman2019}; for Serpens Main and Serpens South from \citet{Herczeg2019}; for RCW~38 from \citet{Getman2019b}; for W~4 from \citet{Cantat-Gaudin18}; and for the remaining regions from \citet{Kuhn19}. Column 5: Number of observed X-ray emitting young stars from \citet{Broos13}. Column 6: Total stellar population down to $0.1$~M$_{\odot}$ inferred from the X-ray luminosity function  (see text).}
\end{deluxetable*}

\section{MYStIX and SFiNCs Datasets} \label{sec:datasets}

Over the past two decades, the {\it Chandra X-ray Observatory} has devoted several months to observations of star forming regions within $d \simeq 3$~kpc of the Sun. {\it Chandra}'s ACIS imager \citep{Garmire03} subtends $17^\prime \times 17^\prime$, and mosaics of multiple pointing are common.  Large projects include a nearly-continuous $\sim 0.9$~Ms exposure of the Orion Nebula Cluster, {\it Chandra} Orion Ultra-deep Project \citep[COUP;][]{Getman05}, $\sim 0.4$~Ms mosaic of the M~17 cluster and environs \citep{Townsley2014}, and a large $1^\circ \times 1^\circ$ mosaic totaling $\sim 1$~Ms of the Carina Nebula complex \citep[CCCP;][]{Townsley11}.  Typical images show hundreds to thousands of faint X-ray sources, most localized to sub-arcsec accuracy.  For each photon, the energy in the range $0.5-8$~keV is recorded with $< 0.3$~keV accuracy, and the arrival time is recorded with $<3.5$~s accuracy.   

Our group has conducted two in-depth analyses of the {\it Chandra} archive of star forming regions observations during the first 10 years of the mission. The MYStIX \citep{Feigelson13} survey covers 20 regions dominated by multiple O stars, most at distances $1.5 \leq d \leq 2.5$~kpc, while the SFiNCs \citep{Getman17} survey covers 22 regions dominated by single O or multiple $\sim$B stars, most at distances $0.3 \leq d \leq 0.8$~kpc.  Several technical papers following \citet{Feigelson13} describe a multistage process of extracting faint X-ray sources from crowded fields, new analysis of UKIDSS near-infrared and {\it Spitzer} mid-infrared data adapted to crowded fields, statistical cross-matching X-ray and infrared catalogs, and applying a naive Bayes classifier to discriminate PMS X-ray stars from extraneous populations such as quasars.  Infrared excess sources and published OB stars were added to the catalog derived from X-ray imagery.

The result of these efforts are catalogs of 31,784 MYStIX and 8,492 SFiNCs probable PMS in the 42 star forming regions \citep{Broos13, Getman17}. Two important datasets are not included in the current super-flare study: the Orion Nebula Cluster COUP campaign where the super-flares are discussed by \citet{Wolk05,Favata2005,Colombo07,Stelzer07,Getman08a,Getman08b}, and the CCCP mosaic of the Carina complex (due to the lack of all necessary X-ray data products). Orion Nebula Cluster flare energy distributions reported in the past studies \citep{Wolk05,Colombo07,Stelzer07} are used as a comparison sample here. Energetics of the brightest modeled COUP flares \citep{Getman08a,Getman08b}, employed as a comparison sample in our companion flare modeling paper (Getman, Feigelson, \& Garmire, 2021, ApJ, submitted), is further discussed in Appendix~\ref{sec:coup_flare_energies}. Thus out of 42 MYStIX and SFiNCs target regions, we study here super-flares from 40 regions that host over 24,000 young stellar objects detected by {\it Chandra}. 

Table \ref{tab:mystix_sfincs_regions} lists these regions, including sky positions, updated $Gaia$-derived distances, numbers of observed X-ray young stellar objects, and estimated total intrinsic stellar populations down to around $0.1$~M$_{\odot}$ following the X-ray luminosity function (XLF) procedures of \citet{Kuhn15a}\footnote{For each of the MYStIX regions, \citet{Kuhn2013a} scale the COUP XLF and \citet{Maschberger2013} IMF to the complete bright parts of the observed XLF and IMF, respectively. The XLF procedures originally applied to the MYStIX regions by \citet{Kuhn15a} are adjusted here to use only two rather than three (as in Kuhn et al.) X-ray energy bands in order to accommodate the lower-counting SFiNCs statistics. All MYStIX+SFiNCs XLFs are also re-calibrated here to the new {\it Gaia}-derived distances.}. 

Our experience from a wide range of MYStIX and SFiNCs studies is that the reliability of these samples is very high, although they are far from complete catalogs of the full Initial Mass Functions of the star forming regions.  They include both PMS stars with and without mid-infrared excess, which we call 'disk-bearing' (Class I and II) and 'diskless' (Class III) PMS stars in this paper. Our previous studies have concentrated mainly on issues characterizing particular star forming regions, and relating to the formation and early evolution of young star clusters.  The present study on X-ray super-flares concentrates on the photon arrival times that played a small role in our previous studies.

\section{Methods} \label{sec:methods}

\subsection{Identification and Classification of X-ray Super-Flares and Mega-Flares} \label{sec:flare_identification}

Our flare selection procedure starts with two quantities tabulated for each {\it Chandra} source by the {\it ACIS Extract} {\it Chandra} and {\it XPHOT} software  packages used to generate the MYStIX and SFiNCs catalogs \citep{Broos10, Getman10}. One is a probability measure of variability within a {\it Chandra} exposure, called ObsID. ObsID durations can range from a few hours to around 2 days early in the mission. {\it ACIS Extract} reports the most significant (among one or more ObsIDs) probability for source variability of the X-ray photon arrival times within an ObsID using a 1-sample Kolmogorov-Smirnov test against a null hypothesis of constant flux.  This measure is called $P_{KS}$.   {\it XPHOT} reports the intrinsic X-ray luminosity of each source averaged over ObsIDs using the local exposure, the distance to the star forming region \citep{Feigelson13, Getman17}, and assuming a typical PMS X-ray spectrum. This measure is called $L_{tc}$ to represent the background-subtracted X-ray luminosity in the {\it Chandra} `total' $0.5-8$~keV band corrected for soft X-ray absorption derived from the observed median energy of extracted photons.  Here we call this quantity $L_X$ or $L_{X,XPHOT}$.  

\begin{figure*}[ht!]
\plotone{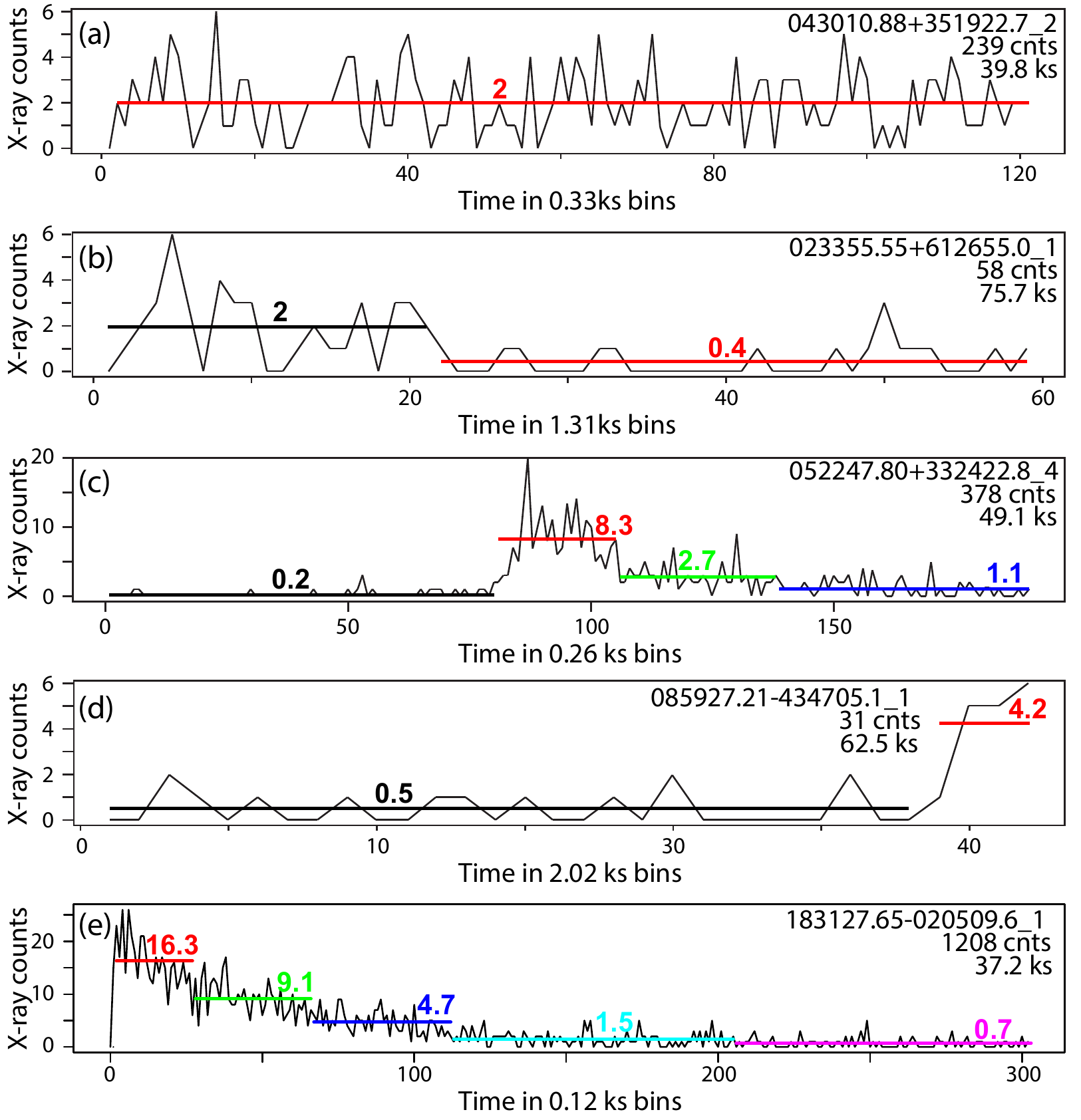}
\caption{Examples of the five classes of segmented lightcurves from the vetting stage: constant (a), variable (b), flare (c), rise (d), and decay(e). Derived segments with averaged counts per bin are colored and labeled. Panel legends list flare names, total numbers of X-ray counts, and arrival time differences between the first and last X-ray counts. The flare name is composed of the flare host star name and relative number of the X-ray {\it Chandra} observation, during which the flare is detected. See Appendix~\ref{sec:change_point_model} for detailed description of such lightcurves. \label{fig:flare_classes}}
\end{figure*}

We then filter the MYStIX and SFiNCs catalogs for the 40 star forming regions with two criteria: $P_{KS} < 0.01$ to locate PMS stars with possible X-ray variability at least in one of the {\it Chandra} ObsIDs, and $\log(L_X) > 30.5$ erg~s$^{-1}$ to locate PMS stars with high time-averaged X-ray luminosities. These selection criteria should capture most luminous flares; rare cases where a brief powerful flare is present in an otherwise faint or undetected source may be missed.  This stage flagged 3,142 ObsIDs from 1,713 stars as luminous and variable X-ray PMS stars.

The second stage seeks to identify the start and stop times of statistically significant variations in the photon arrival times.  In the parlance of time series analysis, this problem  might be called `change point analysis for an inhomogeneous Poisson process' or a 'Poisson regression model with multiple changepoints'.  This problem is well-adapted to likelihood-based fitting if a parametric model for the flare is chosen.  While we could adopt an astrophysically motivated model like `fast rise exponential decay', we choose instead a more general model of a sequence of stepwise constant flux values. This is the approach in the Bayesian Blocks procedure that is widely used to identify and characterize flaring behavior in X-ray sources \citep{Scargle98, Scargle13}.  Our statistical fitting procedure is adopted from methods previously used to address problems in the social sciences \citep{Chib98, Fruhwirth06, Park10, Brandt09}.  It is similar to Bayesian Blocks but gives some additional flexibility.  Our changepoint model and its application to the MYStIX and SFiNCs X-ray light curves are described in Appendix~\ref{sec:change_point_model}.

\begin{deluxetable*}{ccccccccccccc}
%\tablenum{2}
\tabletypesize{\scriptsize}
\tablecaption{MYStIX and SFiNCs F+R+D Flare Host Star Properties\label{tab:flare_host_properties}}
\tablewidth{0pt}
\tablehead{
\colhead{Reg.} & \colhead{Src.} & \colhead{R.A.} &
\colhead{Decl.} & \colhead{$ME$}& \colhead{$\log(L_{X})$} & \colhead{$A_{V}$} & \colhead{$t$} & \colhead{$\log(T_{eff})$} & \colhead{$\log(L_{bol})$} & \colhead{$M$} & \colhead{$R$} & \colhead{$\alpha_{IRAC}$}\\
\colhead{} & \colhead{} & \colhead{(deg)} & \colhead{(deg)} & \colhead{(keV)} & \colhead{(erg/s)} & \colhead{(mag)} & \colhead{(Myr)} & \colhead{(K)} & \colhead{($L_{\odot}$)} & \colhead{($M_{\odot}$)} & \colhead{($R_{\odot}$)} & \colhead{}\\
\colhead{(1)} & \colhead{(2)} & \colhead{(3)} & \colhead{(4)} & \colhead{(5)} & \colhead{(6)} & \colhead{(7)} & \colhead{(8)} & \colhead{(9)} & \colhead{(10)} & \colhead{(11)} & \colhead{(12)} & \colhead{(13)}
}
%\decimalcolnumbers
\startdata
Be 59 & 000053.45+672615.0 & 0.222725 & 67.437501 & 2.4 & 31.0 & 7.0 & 1.7 & 3.63 & -0.03 & 0.8 & 1.8 & -2.2\\
Be 59 & 000054.01+672119.8 & 0.225079 & 67.355504 & 2.4 & 31.0 & 5.5 & 1.0 & 3.56 & -0.27 & 0.5 & 1.8 & -2.0\\
Be 59 & 000102.52+672841.0 & 0.260534 & 67.478076 & 1.9 & 31.0 & \nodata & \nodata & \nodata & \nodata & \nodata & \nodata & -2.2\\
Be 59 & 000138.66+672800.6 & 0.411086 & 67.466849 & 2.0 & 31.0 & 3.3 & 0.7 & 3.52 & -0.41 & 0.2 & 1.9 & -0.5\\
Be 59 & 000144.25+672457.3 & 0.434385 & 67.415918 & 2.5 & 32.1 & 6.2 & 1.7 & 3.75 & 1.19 & 2.7 & 4.1 & -2.2\\
\enddata
\tablecomments{This table is available in its entirety (1027 F+R+D flare host stars) in the machine-readable form in the online journal. A portion is shown here for guidance regarding its form and content. Column 1: Star forming region. Column 2: Source name. Columns 3-4: Source's position for epoch J2000.0 in degrees. Columns 5-6: Source's X-ray median energy and intrinsic X-ray luminosity averaged across all available {\it Chandra} observations using the {\it XPHOT} procedure. Columns 7-13: Source's visual extinction, age, effective temperature, bolometric luminosity, mass, radius, and SED slope as derived through procedures described in \S \ref{sec:stellar_props_appendix}. Stellar properties for five stars (180416.78-242837.2, 180452.94-242705.5, 182001.70-160529.2, 052309.48+332425.2, 085843.86-472856.8), with degeneracy in X-ray-NIR-derived $T_{eff}$ and $L_{bol}$, are obtained through the VOSA SED fitting.  For the remaining stars, the stellar properties are obtained using the methods based on the X-ray and NIR photometry (Appendix \ref{sec:stellar_props_appendix}).}
\end{deluxetable*}

Flare vetting is performed in the third stage of the analyses. Graphical output from the changepoint model method, similar to Figure~\ref{fig:change_points} in Appendix \ref{sec:change_point_model}, combined with observed photon arrival diagrams (\S \ref{sec:flare_atlas}) for the 3,142 ObsIDs, is visually examined and classified into five types as illustrated in Figure~\ref{fig:flare_classes}:
\begin{description}

\item[Constant] Many of the ObsIDs are consistent with a single flux level; that is, the best-fit model has no changepoints. An example is shown in Figure~\ref{fig:flare_classes}a.  This commonly occurs when several ObsIDs are present for a MYStIX/SFiNCs star with statistically significant $P_{KS}$ in one exposure but not all ObsIDs.  Sometimes a spurious changepoint is found in the first few bins due to an unphysical assumption that the light curve starts with zero counts. Of the 3,142 ObsIDs examined, 1,222 constant events are identified.

\item[Variable]  Often the Bayesian segmentation results in two or three flux levels without the appearance of a distinct flare (Figure~\ref{fig:flare_classes}b).  The levels could be different by factors $<1.5$ to factor up to $\sim 10$.  Many cases are probably flares with too few counts to be clearly delineated. But other cases consist of two different, but constant levels of X-ray emission; these may represent the emergence or disappearance of active regions due to stellar rotation. We place  834 variable lightcurves in this category.  

\item[Flare (F)] This class requires at least three segments: a low level before the event, one or more higher levels showing rise and decay, and a low level after the event (Figure~\ref{fig:flare_classes}c).  In most cases, a factor of $\geq 3$ difference between minimum and maximum levels is present. These cases are most useful for science analysis.  Peak X-ray luminosities, durations and total energies can be directly measured, and flare occurrence rates readily calculated.  When more than $\geq 1000$ counts are present in the flare, our companion study gives detailed spectro-temporal modeling of plasma heating and cooling assuming a single-loop geometry (Getman, Feigelson, \& Garmire, 2021, ApJ, submitted). A total of 648 flare events are identified.  

\item[Rise (R)] These are events where sudden onset is seen but the full development of the flare is truncated by the end of the ObsID exposure (Figure~\ref{fig:flare_classes}d). The true flare peak X-ray luminosity may or may not be the observed maximum luminosity.  In most cases, a factor of $\geq 3$ difference between minimum and maximum levels is present.  These cases can be used for flare occurrence rates but have lower limits to their peak luminosities, total energies, and durations. We find 289 rise events among the 3,142 ObsIDs examined.

\item[Decay (D)] There are events similar to the Rise category, but with the beginning of the flare occurring before the beginning of the exposure(Figure~\ref{fig:flare_classes}e).  The maximum luminosity is seen when the observation starts and it typically shows a slow decline in brightness to a constant low level. These cases are treated similarly to the Rise class with lower limits to flare properties. We find 149 decay events.

\end{description} 

In a small fraction of light curves, the flux rises and falls more than once and our 5-component Bayesian segmentation (Appendix~\ref{sec:change_point_model}) is too simple to model the behavior.  These cases are examined individually and flares are manually extracted and placed into the appropriate F, R, or D category. \newline

The flare sample studied here thus consists of 1,086 events: 648 in category F; 289 in R; 149 in D.

\subsection{Properties of Host Stars} \label{sec:host_properties}

Since vast majority of the MYStIX/SFiNCs young stellar objects lack spectroscopic measurements, crude estimates of their source extinctions (in visual band, $A_V$), ages, effective temperatures ($T_{eff}$), bolometric luminosities ($L_{bol}$), radii ($R$), and masses ($M$) are obtained in Appendix \ref{sec:stellar_props_appendix} using {\it Chandra} X-ray and 2MASS, UKIDSS near-infrared (NIR) photometric data described in \citet{Getman14a}, \citet{Kuhn15a} and \citet{Richert18} complemented by fitting  optical-IR spectral energy distributions (SEDs) with the VO SED Anlazyser \citep{Bayo2008} using other additional numerous available optical and IR photometric catalogs. The presence or absence of circumstellar disks is acquired using the {\it Spitzer}-IRAC mid-infrared (MIR) photometry provided in \citet{Kuhn13b}, \citet{Povich13} and \citet{Getman17}. Related methods are described in Appendix \S \ref{sec:stellar_props_appendix}. Table~\ref{tab:flare_host_properties} lists the properties of all 1027 unique F+R+D flare host stars.

\subsection{Flare Properties} \label{sec:energy_estimates}

Three quantities are calculated for each flare (F), rise (R), and decay (D) event: flare duration, $t_{dur}$;  peak luminosity in the $0.5-8$~keV band corrected for soft X-ray absorption, $\log L_{X,pk}$; and total energy, $\log E_X$. The time between the changepoints of elevated emission (as an output from the changepoint model) is used as an initial value for flare duration. Guided by our flare duration choices for the COUP flares \citep{Getman08a}, upon visual inspection of the observed MYStIX/SFiNCs photon arrival diagrams and lightcurves (\S \ref{sec:flare_atlas}), we enlarge these statistical duration times, typically by factors $\times [1.1-2.0]$, in order to better capture the rise and decay tails of the flares. The $L_{X,pk}$ and $E_X$ properties require a conversion from observed counts to X-ray flux, which is then scaled to luminosity and energy using the distance to the host star forming region.  

First, the photon arrival times are converted to count rates ($CR$) with a moving median filter where the bandwidth adaptively varies to include 5 to 500 counts for flares over the range  $N_{phot} < 50$ to $N_{phot} > 5000$ X-ray photons. 

Second, the smoothed $CR$ time series is  converted to X-ray flux using conversion factors based on the median energy ($ME$) and luminosity-dependent spectrum of PMS stars as described by \citet{Getman10} as implemented in the publicly released $XPHOT$ software package.  For most datasets that have $N_{phot}<500$, a single conversion factor is used for the full lightcurve.  For the brighest sources with $N_{phot} > 500$, the conversion factor is calculated at each time in the $CR$ lightcurve based on the local $ME$.  This approach was used by \citet{Getman08a} for the Orion super-flares where the count rate was always high.   Tests on the brightest flares show that the approach based on a single $ME$  gives luminosities about 0.05~dex brighter than the local $ME$ estimator.  This small offset is ignored here. 

Peak $L_{X,pk}$ values are obtained as the maximum in the $L_X$ smoothed lightcurve within the start and stop changepoints  of the flare acquired in \S \ref{sec:flare_identification}.   Total flare energies $E_X$ integrate the $L_X$ values over the flare duration.

The above method of the conversion from the observed X-ray count rates and median energies to intrinsic X-ray luminosities was originally employed in the modeling of the COUP flares by \citet{Getman08a}. This method neglects the contribution of the ``quiescent'' background X-ray emission or ``characteristic'' emission \citep{Wolk05,Caramazza07}, that is probably the product of numerous superposed micro-flares and nano-flares. In their Appendix A, \citet{Getman08a} show that the contribution of the characteristic component to the emission of X-ray super-flares is negligible. It is important to note that the inspection of the electronic atlas of the COUP super-flares \citep[Figure Set 2 in][]{Getman08a} and start/stop times for the COUP flare and characteristic emission segments, listed in Table~1 of Getman et al., suggests that many PMS super-flares are immediately preceded/followed by an elevated X-ray emission for $\sim 4$~days. No correlation is seen between the duration of such elevated emission and stellar rotation period or radius; it is unclear if such emission is associated with the super-flare-host active region or other region(s). Subtraction of such level of emission from our MYStIX/SFiNCs super-flare emission is not astrophysically justified. In the current study we employ super-flare energetics with no subtraction of such elevated emission. Such subtraction will systematically reduce the inferred MYStIX/SFiNCs super-flare energies by around 20\%.

%%%between the first and last changepoints. 

\begin{figure*}[ht!]
\plotone{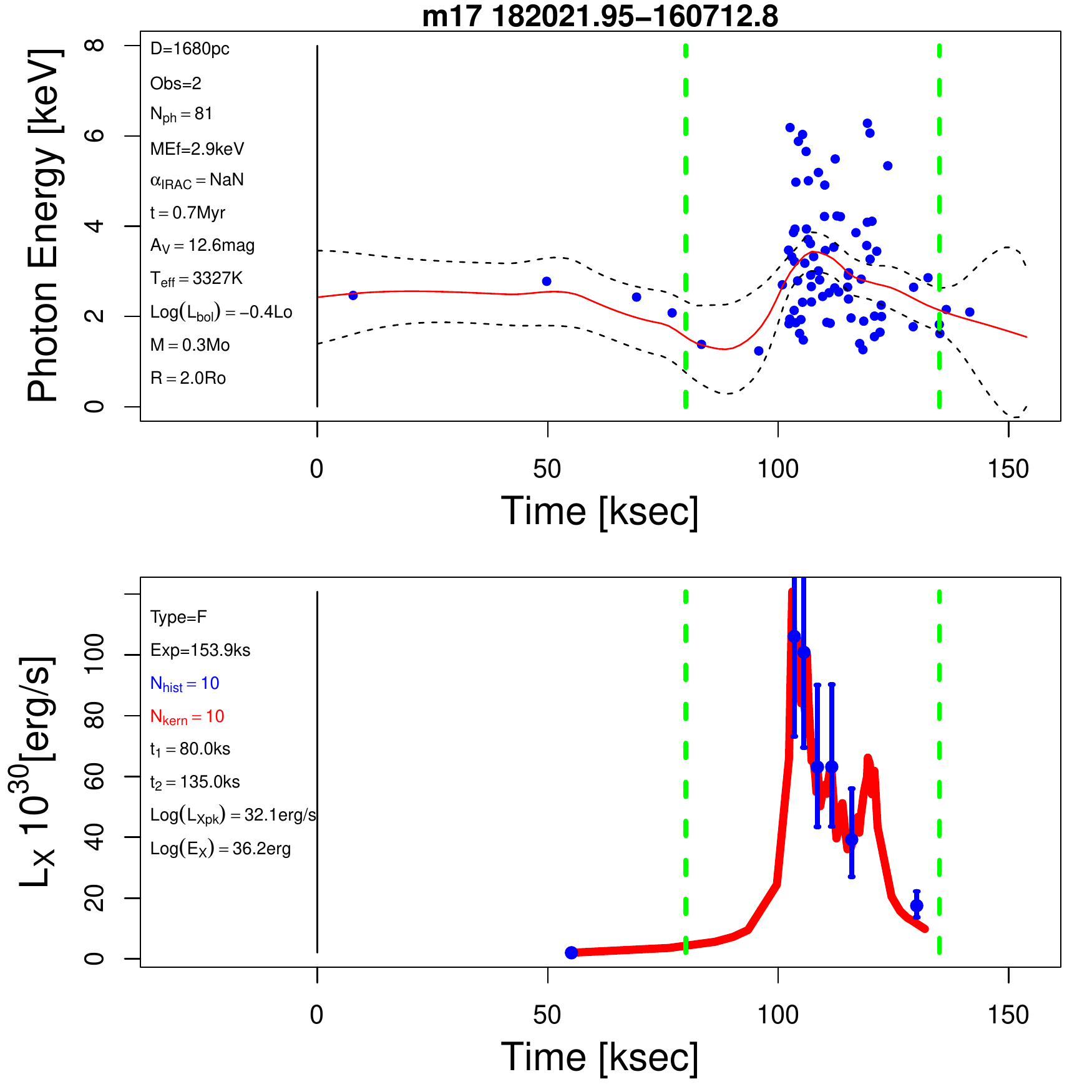}
\caption{Sample page from the flare atlas. See text for details. The complete figure set (1086 images) is available as supplementary material. \label{fig:atlas_example}}
\end{figure*}
\subsection{Treatment of Rise and Decay Flares}

A systematic bias is present for the R and D classes where the observed durations and flare energies are lower limits to the true values.  It is not known whether the observed peak luminosities are correct or similarly biased; we consider them to be lower limits.  In statistics such data are denoted as right-censored data points and the methods from `survival analysis' are used to treat the bias \citep{Feigelson85}.  For univariate samples that include both measured (F class) and lower limits (R and D classes), the nonparametric Kaplan-Meier (KM) estimator provides a corrected distribution function.  \citep{KaplanMeier58}. The KM estimator redistributes the luminosity and energy lower limits to higher values in a maximum likelihood procedure.  KM calculations are made with the {\it survfit} function from the R package {\it survival} \citep{Therneau00,Therneau20}.

\section{Super- and Mega-flare Properties} \label{sec:results}

\subsection{Flare Table and Atlas} \label{sec:flare_atlas}

Identifiers and properties of the 1086 F, R, and D flare events  are provided in Table~\ref{tab:super-flares}. F flares have $19 < N_{phot} < 4,202$~cts, $0.9 < ME < 5.7$~keV, $3 < t_{dur} < 154$~ks, $30.7 < \log L_{X,pk} < 33.8$ erg s$^{-1}$, and $34.3 < \log E_X < 37.6$ erg.  These events are thus `super-flares' and `mega-flares' using the criteria $\log L_{X,pk} \gtrsim  30.5$ erg s$^{-1}$ and $\log E_X \gtrsim  34$ erg suggested in \S\ref{sec:intro}. There are 636 and 450 `super-flares' and `mega-flares' in our sample, respectively (\S \ref{sec:energy_distributions}).

\begin{deluxetable*}{ccccccrcrrr}
%\tablenum{1}
\tablecaption{MYStIX and SFiNCs F+R+D Super-flares\label{tab:super-flares}}
\tabletypesize{\small}
\tablewidth{0pt}
\tablehead{ 
\multicolumn{4}{c}{Star} 
&& \multicolumn{6}{c}{Flare}\\ \cline{1-4} \cline{6-11}
\colhead{Reg.} & \colhead{Src.} & \colhead{ObsID} & \colhead{Expo.} && \colhead{Class} & \colhead{$N_{phot}$} & \colhead{$ME_{f}$} & \colhead{Dur} & \colhead{$\log(L_{X,pk})$} & \colhead{$\log(E_{X})$} \\
\colhead{} & \colhead{} &  \colhead{} &
\colhead{(ks)} && 
\colhead{} & \colhead{(cnts)} & \colhead{(keV)} & \colhead{(ks)} & \colhead{(erg s$^{-1}$)} & \colhead{(erg)}\\ 
\colhead{(1)} & \colhead{(2)} & \colhead{(3)} & \colhead{(4)} &&
\colhead{(5)} & \colhead{(6)} & \colhead{(7)} & \colhead{(8)} & \colhead{(9)} & \colhead{(10)}
}
\startdata
Cep B & 225351.35+623518.3 & 2 & 27.1 & & R & 20 & 2.5 & $>$9.6~~ & $>$31.4~~~~~ & $>$34.9~~~~\\
Cep B & 225355.16+624337.0 & 1 & 26.1 & & F & 1179 & 3.2 & 25.1~~ & 33.0~~~~~ & 37.1~~~~\\
Cep B & 225355.16+624337.0 & 2 & 27.1 & & F & 1003 & 3.3 & 19.0~~ & 32.9~~~~~ & 37.0~~~~\\
Cep B & 225356.68+623436.4 & 3 & 24.1 & & F & 82 & 2.0 & 10.0~~ & 32.1~~~~~ & 35.3~~~~\\
Cep B & 225357.57+622903.6 & 3 & 24.1 & & F & 49 & 2.4 & 15.1~~ & 31.7~~~~~ & 35.2~~~~\\
Cep B & 225414.01+623805.9 & 3 & 27.1 & & R & 105 & 2.3 & $>$12.1~~ & $>$32.0~~~~~ & $>$35.6~~~~\\
\enddata

\tablecomments{Only a few examples of the table entries are given here; the full machine readable table for all 1086 F+R+D flares is provided in the electronic edition of this paper. Column 1: Star forming region. Column 2: Flare host star name. Column 3: The relative number of the X-ray {\it Chandra} observation, during which the flare is detected. Complete lists of the MYStIX and SFiNCs {\it Chandra} observations with full ObsID names are available in \citet{Kuhn2013a,Townsley2014,Getman17}. Column 4: {\it Chandra} ObsID exposure time in kilo seconds. Column 5: Flare class. Columns 6-10: Flare properties:  X-ray counts, flare median photon energy, flare duration, peak X-ray luminosity, and flare energy.}
\end{deluxetable*}

We provide an atlas in which both tabulated and graphical information on each of the 1086 F, R and D flare events are collected onto a single page.  A  sample atlas page is shown in Figure~\ref{fig:atlas_example}.  This is an X-ray mega-flare from a low-mass PMS star embedded in the M17 North Bar cloud \citep{Broos07}. The page features two plots with the host star and flare properties extracted from Tables~\ref{tab:flare_host_properties} and \ref{tab:super-flares}.  The first plot gives X-ray photon arrival times and energies with individual photons marked as blue points.  The red curve shows a likelihood-based local quadratic regression fit with 84\% confidence intervals (black dashed lines) generated using the {\it locfit.robust} function from CRAN package {\rm locfit} package \citep{Loader20}. This procedure and its mathematical foundations are described by \citet{Loader99}.  The flare changepoints derived in Appendix~\ref{sec:change_point_model} are indicated by the green dashed lines.  Time ranges from the start to the end of the {\it Chandra} ObsID exposure. 

The plot title gives the star identifier from the MYStIX or SFiNCs catalog and its star formation region. The annotation gives 11 scalar quantities: distance to the star formation region; relative number of the current ObsID for this X-ray source; number of photons in the ObsID with the flare; median energy of the ObsID; infrared slope from which disk presence is inferred; estimated age, visual absorption, stellar effective temperature, bolometric luminosity, mass and radius derived using optical-IR photometry data as described in \S \ref{sec:stellar_props_appendix}.  

The second plot gives an adaptively smoothed absorption-corrected X-ray luminosity lightcurve (red) with 1-$\sigma$ confidence intervals for a binned histogram (blue) using analytical approximations for a Poisson distribution \citep{Gehrels86}. The binned histogram is composed of independent count bins, each accumulating similar numbers of X-ray counts ($N_{hist}$) and centered at the mean arrival time between the first and the last counts in the bin. The plot legends include: flare type (\S\ref{sec:flare_identification});  ObsID exposure time; number of counts per adaptive kernel (red curve); number of counts in a histogram bin (blue); flare start and stop times (green lines), which are the adjusted changepoints as described in \S \ref{sec:energy_estimates}; $\log L_{X,pk}$; and $\log E_X$.  On the X-ray luminosity time series plots, not all jiggles in the red curve are statistically significant, and blue circles-with-errors are not carefully placed with respect to possible interesting structures such as flare peaks.

Two idiosyncracies of this particular star can be noted.  First, the {\it Spitzer}-IRAC counterpart is missing so no $\alpha_{IRAC}$ value is given because it lies near the cloud--H~$_{\rm{II}}$ interface which suffers from bright MIR nebula emission. Second, the source remains undetected in X-rays within the first half of the $Chandra$ exposure prior to the flare.

\begin{figure*}[ht!]
\plotone{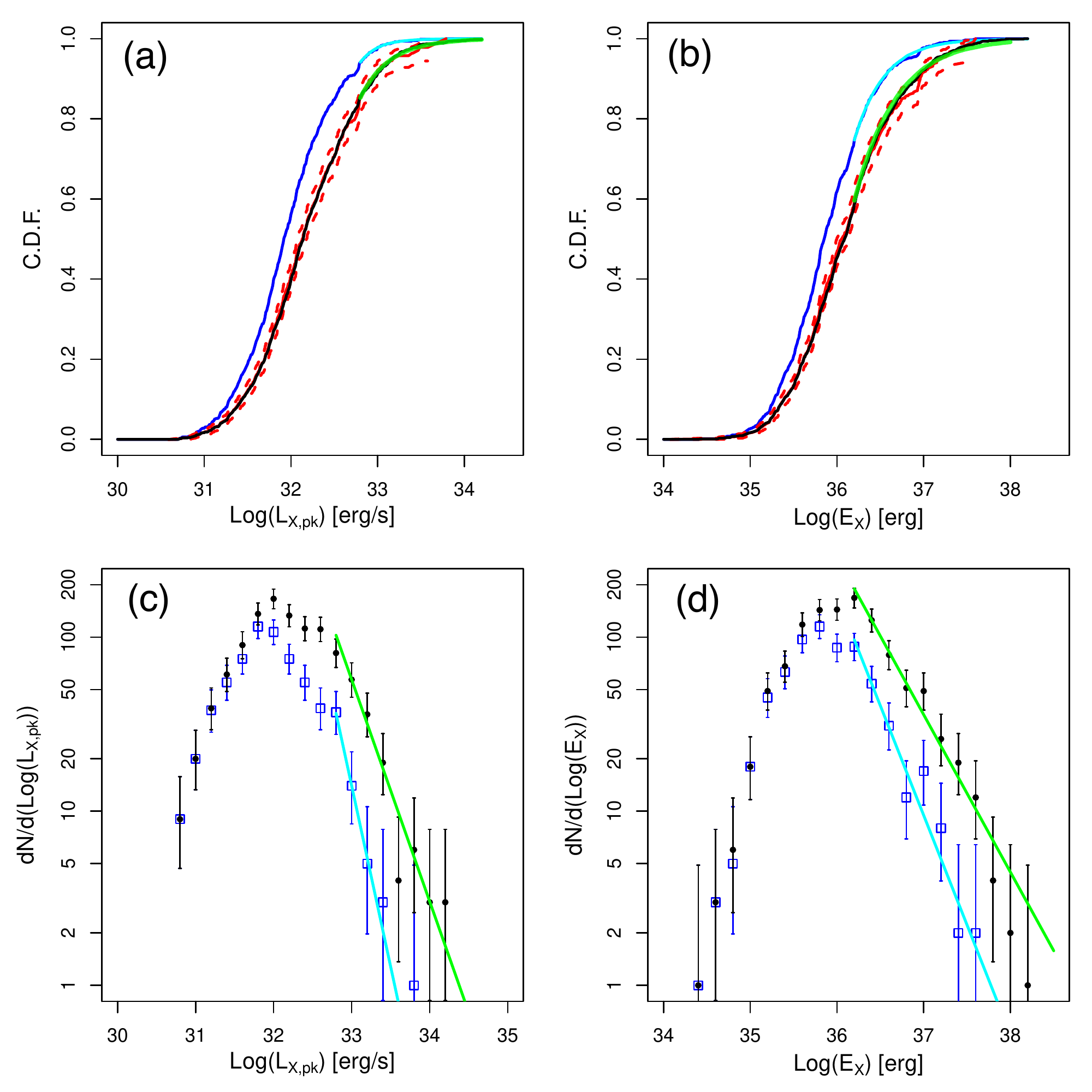}
\caption{Cumulative distribution functions (top panels) and corresponding differential histograms with 0.2 dex bins (bottom panels) of the peak flare X-ray luminosities $L_{X,pk}$ (left) and flare energies $E_X$ (right). The completely observed 648 F flares are shown in blue, and the Kaplan-Meier estimators for the full sample of the 1086 F+R+D flares are shown in red (solid red) with 95\% confidence bands (dashed red). The full sample of 1086 F+R+D flares, for which the energies and peak X-ray luminosities of the ‘R,D’ flares are multiplied by 5 to match the KM estimator, is shown in black (lines in (a,b) and points in (c,d)).} Pareto (power law) fits to the high luminosity/energy tails are shown in cyan and green for the F and F+R+D samples, respectively. Histogram error bars are approximate 95\% confidence intervals. \label{fig:cdf_kaplan_meier_full}
\end{figure*}

\subsection{Flare X-ray Peak Luminosity and Energy Distributions} \label{sec:energy_distributions}

PMS flaring has been most intensively studied in the Orion Nebula Cluster for two reasons: the photon flux for flares is high due to its close distance of only $0.4$~kpc, and the COUP project provides a unique almost-continuous {\it Chandra} observation over 13 days \citep{Getman05}.  Studying a sample of $\sim 1$~M$_\odot$ stars, \citet{Wolk05} reported the flare energy distribution slope of $dN/dE_{X} \propto E_X^{-1.7}$ using a linear regression technique, which was revised to $dN/dE_{X} \propto E_X^{-1.9 \pm 0.2 }$ by \citet{Stelzer07} using a maximum likelihood procedure.  \citet{Caramazza07} found an X-ray flare count ($C_X$) distribution of $dN/dC_X \propto C_X^{-2.2}$ for 151 flares from low-mass $0.1-0.3$~M$_{\odot}$ COUP stars. For the high energy tail of a larger sample of 954 COUP flares regardless of host stellar mass, \citet{Colombo07} finds the slope of $E_X^{-2.1}$.

In addition to the Orion Nebula Cluster,  an important survey of the Taurus molecular cloud was made with the {\it XMM-Newton} satellite \citep[XEST;][]{Gudel07}.  The energy distribution of 33 X-ray flares was found to be $dN/dE_{X} \propto E_X^{-2.4 \pm 0.5}$ \citep{Stelzer07}.

Figures \ref{fig:cdf_kaplan_meier_full}a-b show the KM estimators of peak luminosities and total energies for the full sample of 1086 MYStIX/SFiNCs flares (red curve), together with the empirical cumulative distribution functions for the fully observed subsample of 648 F class flares (blue curve). The R and D flares raise the median values from $\log L_{X,pk}=31.9$ to 32.2 erg~s$^{-1}$ and from $\log E_X=35.8$ to $36.1$ erg. Quite reasonably, the R and D flares that are  truncated by the limited duration of {\it Chandra} exposures tend to be more luminous and more energetic than those captured in the entirety as F flares.  This is consistent with the finding of \citet{Colombo07} that the flares identified within the shorter 100~ksec COUP {\it Chandra} blocks appear systematically less energetic than those identified within the entire COUP observation. The black lines in Figures \ref{fig:cdf_kaplan_meier_full}a-b and black points in Figures \ref{fig:cdf_kaplan_meier_full}c-d represent the full sample of 1086 F+R+D flares, for which the energies and peak X-ray luminosities of the `R,D' flares are multiplied by 5 to match the KM estimator (red). The unbinned cumulative distribution functions for this corrected by $\times 5$ F+R+D sample (i.e., the C.D.F. black lines in Figures \ref{fig:cdf_kaplan_meier_full}a-b) are fitted with the Pareto function to obtain the Pareto slope $\beta$, as detailed below.  

Both flares from individual regions and from a collection of regions (like MYStIX/SFiNCs) may be subject to spatially varying absorption across a region (\S \ref{sec:absorption_effects}). Unlike the flare luminosity/energy distributions for individual star forming regions, the MYStIX/SFiNCs distributions may be subject to the additional effect of different distances (thus source/flare sensitivities) towards different MYStIX/SFiNCs regions. This is not a problem for the complete mega-flare sample (\S \ref{sec:distance_effects}). 

Figures \ref{fig:cdf_kaplan_meier_full}c-d show  differential distributions of the upper panel c.d.f.'s grouped in 0.2 dex bins.  Histogram error bars are approximations to 95\% confidence intervals of a Poissonian distribution \citep{Gehrels86}. The lower bins clearly represent incomplete sampling as our super-flare selection procedure (\S\ref{sec:flare_identification}) requires that the time-averaged luminosity exceeds $\log L_X = 30.5$ erg~s$^{-1}$.  Many flares with $31 < \log L_{X,pk} < 32$ erg~s$^{-1}$ will be diluted by long periods of non-flaring emission so the time-averaged luminosity falls below our selection limit.  Notice that these histogram representations of the data are shown here to visually emphasize approximate data completness limits. Based on the peak values in these histograms, our sample appears complete above the peak of the differential distributions at $\log L_{X,pk} > 32.0 $ erg~s$^{-1}$ and $\log E_X \simeq 36.0$ erg. More accurate completeness limits are derived below based on the Pareto fits to the unbinned data (i.e. those shown as C.D.F.s in Figures \ref{fig:cdf_kaplan_meier_full}a-b). 

It is well-known that solar and stellar flares exhibit power law (Pareto function) distributions of various properties. Choosing progressively higher X-ray luminosity and energy cut-offs following \citet{Stelzer07}, the unbinned cumulative distribution functions (c.d.f.'s) with $N=1086$ or less data points are fitted by maximum likelihood estimation to the Pareto distribution function   
\begin{equation} \label{eq:Pareto}
C.D.F.~  = ~1-(x_{min}/x)^{\beta} ~~{\rm for}~ x \geq x_{min} ~~ {\rm where} \nonumber
\end{equation}
\begin{equation}
\beta ~  =  ~ \frac{N}{\sum_1^N ln(x/x_{min})}.
\end{equation}
The energy (or X-ray luminosity) distributions are expressed through the Pareto slope $\beta$ as $\log(dN/d\log(E_{X})) \propto \log(E_{X})^{-\beta}$ or $dN/dE_X \propto E_X^{-\beta - 1}$. 

For the entire F+R+D flare sample, the Anderson-Darling goodness-of-fit test shows statistically unacceptable fits with $p < 0.01$ at the energy cut-offs $\log(E_X) < 36.1$~erg, consistent with the shape of the differential distribution (Figure \ref{fig:cdf_kaplan_meier_full}d). The fits above that energy value show acceptable and statistically indistinguishable ($p>0.05$) solutions with the powerlaw slope varying between $\beta = [0.91 - 1.00]$ for the energy cut-offs $\log(E_X) = [36.1-36.5]$~erg, where the data samples are the richest, $N> [200-500]$ data points. Conservatively, we choose the completeness limit as $\log(E_X) = 36.2$~erg. This completeness limiy is the reason we choose $\log(E_X)=36.2$~erg as the threshold for the label `mega-flare' in contrast to `super-flare'.  

With this energy cut-off value, the energy distribution has a powerlaw slope $\beta = 0.95$. Slope uncertainties (95\% confidence intervals) obtained from 1000 bootstrap resamples for the KM estimators are $\pm 0.07$.  The Pareto model with $\beta = 0.95 \pm 0.07$ above $\log(E_X) = 36.2$~erg is shown as the green curve in Figure \ref{fig:cdf_kaplan_meier_full}b.

At higher energy cut-offs $\log(E_X)=[36.8-37]$~erg, the powerlaw slope changes to higher values of $\beta=[1.2-1.4]$, but with fewer sample data points $N=[80-140]$ and hence higher statistical uncertainties, $\pm 0.2$. The outlier points, visually represented by the binned point at $\log(E_X) = 37$~erg (Figure \ref{fig:cdf_kaplan_meier_full}d), are  likely the cause of this slope increase. 

For the smaller `F' flare sample, the Pareto slope is $\beta = 1.27 \pm 0.16$ at $\log(E_X) = 36.2$~erg (the cyan lines in Figures \ref{fig:cdf_kaplan_meier_full}b and d.

The inferred Pareto slope of $\beta = 0.95 \pm 0.07$ for the F+R+D sample leads to the X-ray flare energy distribution of $dN/dE_X \propto E_X^{-\beta - 1} =  E_X^{-1.95}$ within the energy range of $\log(E_X)=36.2$ to $38$ erg. This $-1.95$ power law energy distribution is consistent with those for optical, EUV, and X-ray solar/stellar flares captured at a very wide but lower range of energies, from $E_{flare} = 10^{24}$~erg for solar nanoflares to $E_{flare} = 10^{35}$~erg for super-flares from Solar-type stars \citep[e.g.,][and references therein]{Notsu19,Okamoto2020}. The MYStIX/SFiNCs flare energy distribution is also consistent with that of the aforementioned super-flare samples from young stars in the Orion Nebula and Taurus star forming regions \citep{Wolk05, Stelzer07, Colombo07, Caramazza07}. Compared to these previous studies, the MYStIX/SFiNCs data offer a factor $>6$ increase in the sample of mega-flares from young stars.

\begin{figure*}[ht!]
\plotone{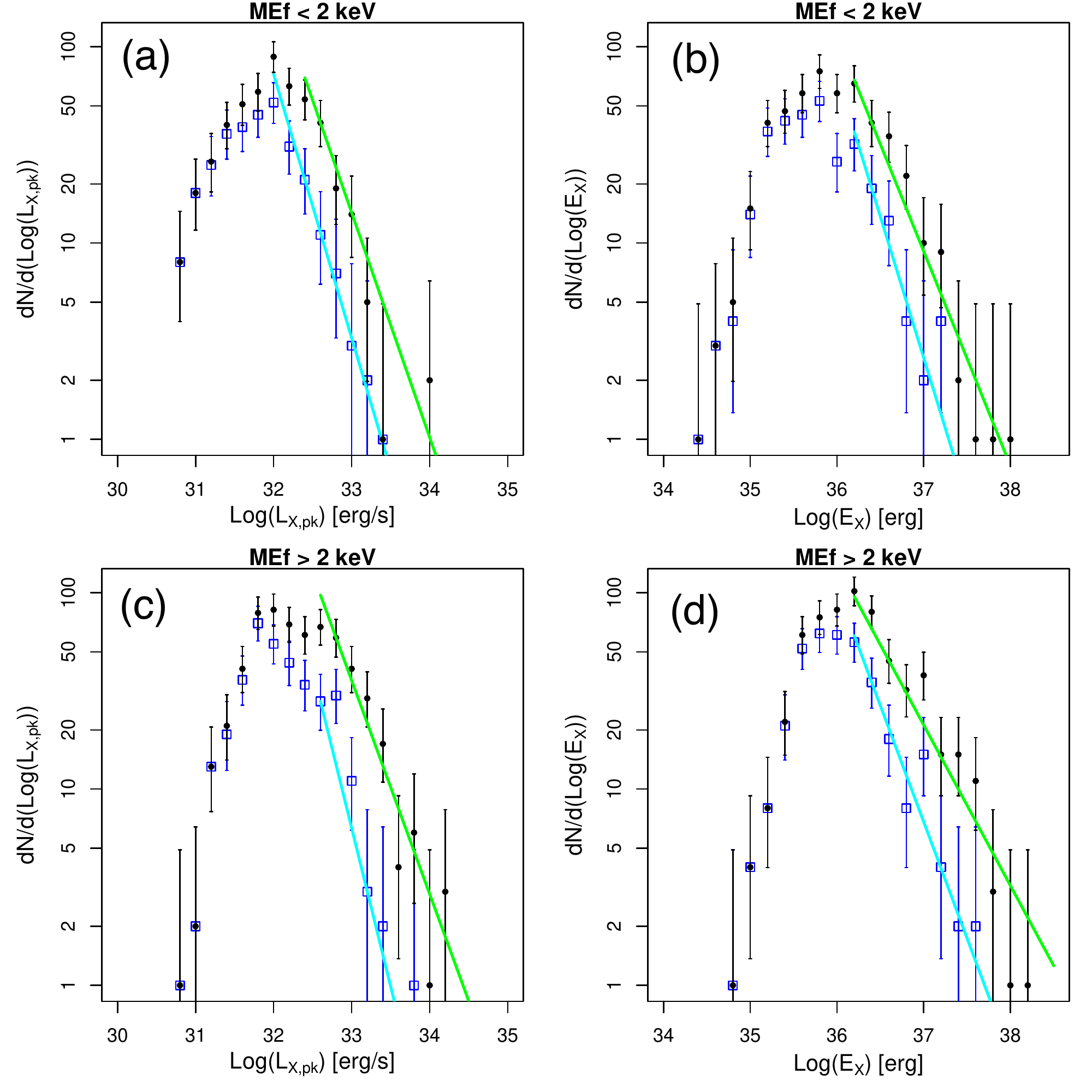}
\caption{As in Figures \ref{fig:cdf_kaplan_meier_full}c-d, differential histograms with 0.2 dex bins, but for flare samples stratified by absorption. The peak X-ray luminosity (left panels) and flare energy (right panels) for the super-flares stratified by low-absorption (top panels) and high-absorption (bottom panels) systems. `F'-flare samples are in blue ($N=299$ unabsorbed and $N=349$ absorbed flares)  and `F+R+D' flare-sample are in black ($N=490$ unabsorbed and $N=596$ absorbed flares). For the `F+R+D' samples the energies and luminosities of `R,D' flares were inflated by a factor of 5 to match the corresponding KM estimators (the KM estimators are not shown). Power law slopes (derived from the Pareto fits to unbinned data based on the approach desribed in \S \ref{sec:energy_distributions}) are shown as green and cyan lines for the `F+R+D' and `F' samples, respectively. \label{fig:KM_histogram_absorbed}}
\end{figure*}

We thus find that the shape of the energy distribution of stellar flares at their highest levels are similar to solar and stellar flares over a remarkable 14 orders of magnitude in energy.

For the $L_{X,pk}$ distribution of the F+R+D flare sample, the Pareto fits to the data become statistically acceptable at $\log(L_X) =32.5$~erg~s$^{-1}$ and onwards. The inferred slopes are $\beta = 1.11 \pm 0.09$ and $\beta = 1.26 \pm 0.16$ at the luminosity cut-offs of $32.5$ and $32.8$~erg~s$^{-1}$, respectively. The latter solution is shown in green in Figures \ref{fig:cdf_kaplan_meier_full}a and c. This progression from a shallower to a bit steeper slope reflects the broken power law morphology of the $L_{X,pk}$ histogram  that is clearly seen in Figure \ref{fig:cdf_kaplan_meier_full}c. The latter is presented in the next section. The distributions for the `F' flare sample have a similar broken powerlaw shape but with systematically steeper slopes (cyan curves in Figure \ref{fig:cdf_kaplan_meier_full}a and c.

\subsection{Effects of Absorption} \label{sec:absorption_effects}

Young stellar objects are often subject to soft X-ray absorption by K- and L-shell transitions in metal atoms along the line of sight \citep{Wilms00}.  Most of this absorption occurs locally in the parental molecular clouds and, for the youngest stars, their local protostellar envelopes.  Absorption is thus an indirect measure of stellar age with the youngest Class I and II systems more absorbed than older Class III systems. But for the more distant star formation regions, absorption by molecular clouds in intervening spiral arms also contributes to absorption.  The line-of-sight material causing soft X-ray absorption  is often quantified as $N_H$, the column density of equivalent hydrogen, which can be converted to a visual absorption $A_V$ by assuming a gas-to-dust ratio \citep[e.g.,][]{Hasenberger2016,Zhu2017}. In the MYStIX and SFiNCs studies, the median energy of X-ray photons $ME$ in keV is used as a surrogate for $N_H$ following the calibration procedure described by \citet{Getman10}.  Here we measure $ME_f$, the median energy of photons arriving between the start and end times of the flares.

Figures \ref{fig:KM_histogram_absorbed} present the histograms of the peak X-ray luminosity and flare energy distributions for lightly ($ME_f \leq 2$~keV) and heavily ($ME_f > 2$~keV) absorbed  flares, respectively. The 2~keV boundary corresponds approximately to $\log N_H \simeq 22.0$ cm$^{-2}$ and $A_V \simeq 5$ mag.  We see that the shape of the $L_{X,pk}$ distribution is especially sensitive to the absorption effect.  The lightly-absorbed flare samples follow a powerlaw distribution reasonably closely over the $\log(L_{X,pk,lim}) \sim 32-34$~erg/s range, but the heavily-absorbed samples show a deficit around $\log(L_{X,pk,lim}) \sim 32.2-32.8$~erg/s.  

The reason for this deficit is again probably related to our selection criteria for selecting $\ga 1000$ superflaring stars from $\ga 24,000$ MYStIX and SFiNCs stars.  We require that the time-averaged X-ray luminosities exceed $\log L_X > 30.5$ erg s$^{-1}$ in the total $Chandra$ $0.5-8$ keV band (\S \ref{sec:flare_identification}).  The recovery of missing soft X-ray emission from heavily absorbed sources appears to be incomplete, and thus a significant number of stars with X-ray flare peak luminosities around $\log L_{X,pk} \simeq 32$ erg s$^{-1}$ are excluded by our time-averaged $\log L_X > 30.5$ erg s$^{-1}$ selection criterion.

The flare energy distributions are less affected by absorption. Based on the Pareto fits to the unbinned data (C.D.F.s are not shown here), both, the low- and high-absorbed flare energy distributions are complete at around $\log(E_X) = [36.1-36.2]$~erg with statistically indistinguishable Pareto slopes for the F+R+D flare samples of $\beta=[1.02-1.09] \pm 0.14$ and $\beta = [0.86-0.89] \pm 0.08$, respectively (green in Figures \ref{fig:KM_histogram_absorbed}b and d).

\subsection{Effects of Distance} \label{sec:distance_effects}
To consider the possible sample bias due to the range of distances to the MYStIX/SFiNCs star forming regions, the F+R+D flare sample is divided into two flare groups, near ($D \le 1500$~pc) and far ($D > 1500$~pc). Application of our analyses to the flare energy distributions of these two groups shows that the nearby flares have lower completeness limits of $\log(E_X) =[35.9-36.0]$~erg with $N=[196-167]$ flares above these limits and corresponding Pareto slope range $\beta=[0.87-0.93] \pm 0.10$. For more distant flares, the completeness limits are similar to those of the entire flare sample: $\log(E_X)=[36.1-36.2]$~erg with $N=[387-333]$ and $\beta=[0.88-0.94] \pm 0.08$. The inferred flare energy Pareto slopes remain statistically indistinguishable between these two distance-stratified samples.    

\section{Super/Mega-flares and Host Star Properties} \label{sec:allmystixsfincs_vs_super-flares}

\subsection{Mega-flares from Protostars}
\label{sec:protostars}

The reports of X-ray flaring in the Orion cloud Class~0 protostar HOPS~383 and other early-phase protostars pushes the onset of X-ray flaring into the earliest infall stages of star formation \citep[][and references therein]{Grosso20}. For this Class~0 protostar, {\it Chandra} detected 28 X-ray counts with median energy  5.4~keV corresponding to the column density of $\log(N_H) \simeq 23.8$~cm$^{-2}$ or $A_V \simeq 200$ mag. Our MYStIX+SFiNCs super-flare sample has eight heavily embedded protostellar candidates with flare X-ray median energies above 5.0~keV.  These sources, which lack mass estimates, are listed in Table~\ref{tab:protostars}.  See the electronic flare atlas (\S \ref{sec:flare_atlas}) for their photon arrival diagrams and lightcurve morphology. Three sources are located in the nearby ($d \sim 300-400$~pc)  NGC~1333, L1251b, and Flame regions; one is in the intermediate-distant ($d \sim 900$~pc) Cep~A region; and four are in the more distant ($d \sim 1700$~pc) M~17 and RCW~38 regions. SED IRAC slopes,  available for the 3 out of 8 sources, show fluxes ascending towards longer wavelengths confirm their protostellar nature. 

For the 5 F (fully observed) MYStIX/SFiNCs protostellar flares in Table~\ref{tab:protostars}, the median X-ray flare peak luminosity of $L_{X,pk}=3\times10^{32}$~erg~s$^{-1}$, a factor of 7 higher than the flare seen in HOPS~383. Using the $\log E_X = 36.2$~erg boundary, at least 5 of the 8 events are mega-flares.  This clearly demonstrates that the extremely high levels of flaring seen in Class II and III PMS stars is present in Class I, and possibly Class 0, protostars. 

Followup far-IR/sub-millimeter observations of these 8 X-ray sources would assist in unravelling their evolutionary stages and looking for spectroscopic evidence that the penetrating super-flare X-rays play a role in ionization of their circumstellar envelopes and disks.

\begin{deluxetable*}{clrcccrcrrr}
%\tablenum{2}
\tablecaption{Super-flares from Protostellar Candidates\label{tab:protostars}}
\tablewidth{0pt}
\tablehead{
\colhead{Region} & \colhead{Source\_Obs} & \colhead{R.A.} &
\colhead{Decl.} & $\alpha_{IRAC}$& \colhead{Flare} & \colhead{$C_f$} & \colhead{$ME_f$} &
\colhead{$\log(L_{X,pk})$} & \colhead{$\log(E_{X})$}\\
\colhead{} & \colhead{} &  \colhead{(deg)} &
\colhead{(deg)} & \colhead{} & \colhead{} & \colhead{(cnts)} & \colhead{(keV)} &
\colhead{(erg/s)} & \colhead{(erg)}\\
\colhead{(1)} & \colhead{(2)} & \colhead{(3)} & \colhead{(4)} & \colhead{(5)} &
\colhead{(6)} & \colhead{(7)} & \colhead{(8)} & \colhead{(9)} & \colhead{(10)}
}
%\decimalcolnumbers
\startdata
CepA & 225619.58+620223.4\_1 & 344.081599 & 62.039843 & 0.3& F & 38~~~ & 5.7 &  32.4~~~~~ & 36.6~~~\\   
Flame & 054143.54-015511.7\_1 & 85.431458 & -1.919931 & 2.1& R & 21~~~ & 5.4 & $>31.3$~~~~~ & $>34.9$~~~\\   
L1251b & 223846.92+751133.6\_2 & 339.695508 & 75.192679 & \nodata& F & 77~~~ & 5.4 &  32.3~~~~~ & 36.0~~~\\ 
M17 & 182016.85-160726.0\_6 & 275.070224 & -16.123908 & \nodata& R & 22~~~ & 5.1 &  $>32.8$~~~~~ & $>36.2$~~~\\   
M17 & 182021.76-161257.6\_2 & 275.090706 & -16.216027 & \nodata& F & 296~~~ & 5.2 &  33.8~~~~~ & 37.6~~~\\  
M17 & 182022.11-161305.2\_2 & 275.092133 & -16.218137 & \nodata& R & 53~~~ & 5.2 &  $>32.8$~~~~~ & $>37.0$~~~\\  
NGC1333 & 032858.43+312217.7\_2 & 52.243473 & 31.371592 & 2.0& F & 67~~~ & 5.4 &  32.0~~~~~ & 35.8~~~\\ 
RCW38 & 085906.63-473021.9\_1 & 134.777644 & -47.506100 & \nodata& F & 244~~~ & 5.0 &  33.5~~~~ & 37.5~~~\\
\enddata
\tablecomments{Column 1: Star forming region. Column 2: Unique X-ray flare name, composed of the X-ray source name and the relative number of the X-ray {\it Chandra} observation, during which the flare is detected. Columns 3-4: The source position for epoch J2000.0 in degrees. Column 5: SED IRAC slope. Column 6: Flare type: F=full, R=Rise, D=Decay. Columns 7-10: Flare properties, including  X-ray counts, median photon energy, peak X-ray luminosity, and flare energy.}
\end{deluxetable*}

\subsection{Comparing Disk-bearing and Diskless Stars} \label{sec:disky_diskless}

As outlined in \S~\ref{sec:intro}, it has been long debated whether some X-ray flares from PMS stars arise from the magnetic reconnection in loops extending from the star to the disk rather than loops with both footprints in the stellar surface.  If we can assume that the near- and mid-infrared photometric excess is an adequate indicator of the presence of gaseous inner protoplanetary disk, we can investigate this issue by comparing flare distributions in MYStIX/SFiNCs stars with and without disks.

\begin{figure}[ht]
\plotone{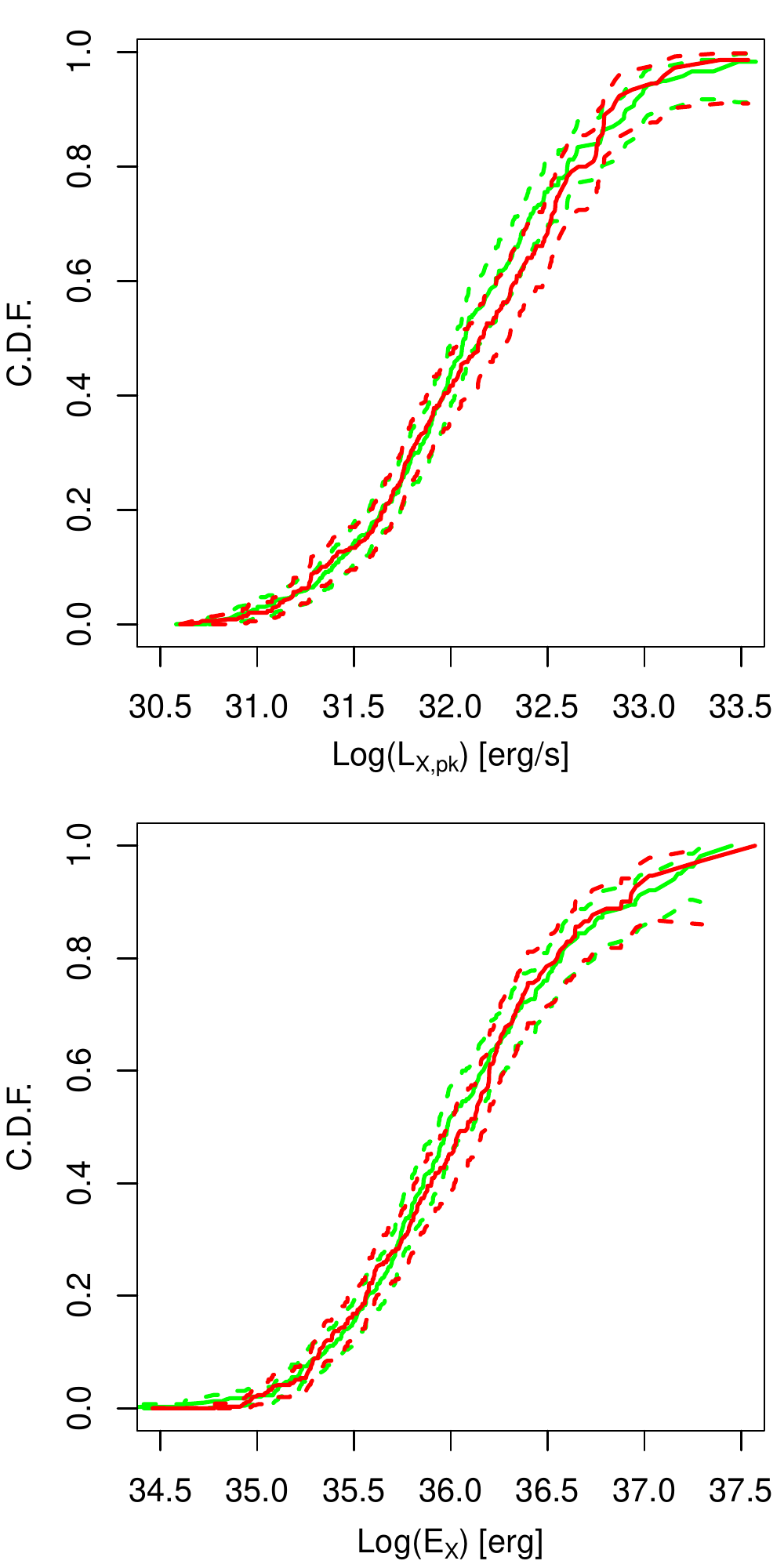}
\caption{Kaplan-Meier cumulative distribution estimators of flare peak X-ray luminosity and flare energy for F, R and D super-flares for disk-bearing (red) and diskless (green) stars. Dashed curves give 95\% confidence intervals.  \label{fig:cdf_disky_vs_diskless}}
\end{figure}

We associate diskless stars with infrared spectral energy distribution slopes $\alpha_{IRAC} \leq -1.9$ and  disk-bearing stars with $\alpha_{IRAC} > -1.9$ \citep{Richert18}.  Two-thirds of the hosts of the 1086 super-flarestars have sufficient infrared photometry to measure $\alpha_{IRAC}$, roughly evenly divided between the two classes with 397 flares from diskless and 348 flares from disk-bearing stars.  Figure \ref{fig:cdf_disky_vs_diskless} compares the KM estimators for peak X-ray luminosity and flare energies for these two subsamples.  The c.d.f. estimators appear indistinguishable.  This is validated with the survival analysis logrank 2-sample test \citep{Harrington82} with p-value $>0.5$ for both measures of flare strength. 

Similarly, if only the most powerful flares with energies above the completeness limit of $\log(E_X) = 36.2$~erg are considered, the flare samples are reduced to 85 and 69 flares from diskless and disk-bearing stars, respectively. The c.d.f. estimators for peak X-ray luminosity and flare energy of the disk-bearing and diskless stars remain indistinguishable, with logrank p-values $>0.5$ (figure is not shown).  

We thus find no statistical differences in flare strength distributions between disk-bearing and diskless MYStIX/SFiNCs samples.  Both types of young stellar objects, with and without disks, produce X-ray super-flares that follow similar distributions of flare peak luminosity and energy. 

Consistent with our result but for much smaller numbers of X-ray flares, \citet{Stelzer07} report no flare energy differences for flares detected in disk-bearing and diskless stellar members of the Taurus star forming region. No differences in the flare occurrence rates and flare durations are seen between disky and dikless COUP stars \citep[\S 5.2 in][]{Flaccomio2012}. Furthermore, no noticeable differences in the relations between the optical and X-ray flare energies are seen for flares detected from disk-bearing and diskless members of the NGC 2264 region \citep{Flaccomio2018}.     

\subsection{Super-flares and Stellar Mass} \label{sec:flare_vs_mass}

\begin{figure*}[ht]
\epsscale{1.2}
\plotone{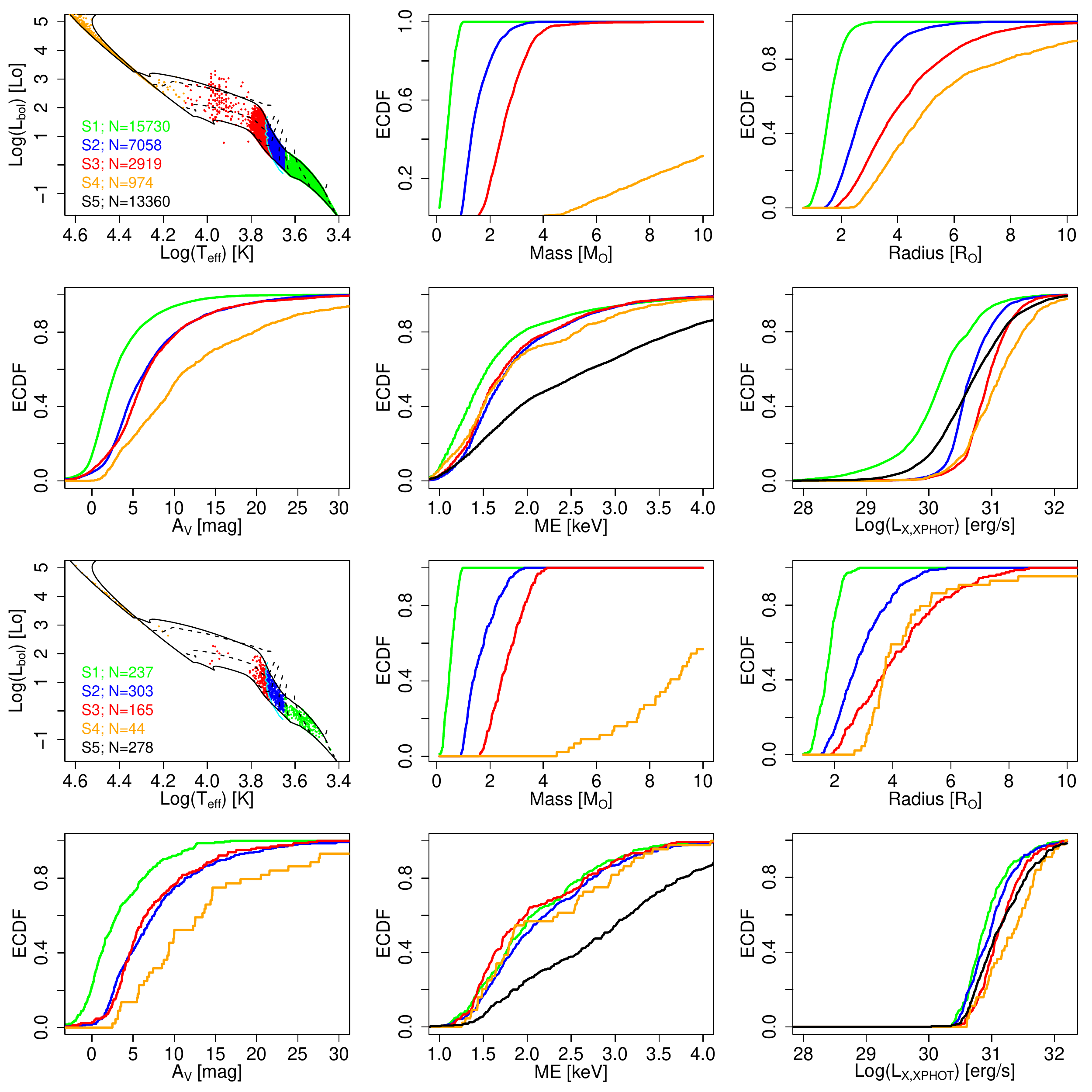}
\caption{Hertzsprung-Russell diagrams and empirical cumulative distribution functions of stellar mass, radius, visual extinction, X-ray median energy, and X-ray luminosity.  The upper six panels include all MYStIX+SFiNCs stars, and the lower six panels are restricted to the super-flare hosts. Color-coded source strata are described in the text. In the HRDs, the black solid curves show theoretical isochrones at 0.4 and 5 Myr, and dashed curves show mass tracks for 0.1, 0.5, 0.8, 1.2, 2, 3 and 5 M$_\odot$. The cyan curve in the HRD between the solar-mass (blue) and intermediate-mass (red) strata denotes the approximate boundary between the Hayashi and Henyey evolutionary tracks. The legends list numbers of young stars in each of the source strata. \label{fig:super-flares_vs_allmystixsfincs}}
\end{figure*}

Our sample has 1027 young stellar objects that produce 1086 X-ray `F+R+D' super- and mega-flares; 749 of these stars have available stellar mass estimates (Table~\ref{tab:flare_host_properties}).  To understand the nature of these  host stars, their properties are compared to the full sample of  MYStIX+SFiNCs young stars with available masses and XPHOT X-ray luminosities. The six properties of interest include: location on the Hertzsprung-Russell diagram (HRD); stellar mass and radius; source visual extinction and source X-ray median energy measuring line-of-sight absorption; and source X-ray luminosity averaged across all available {\it Chandra} observations ($L_{X,XPHOT}$). Figure~\ref{fig:super-flares_vs_allmystixsfincs} compares these six properties for all MYStIX+SFiNCs stars (upper panels) and super/mega-flare hosts only (lower panels). The derivation of the stellar properties is detailed in Appendix~\ref{sec:stellar_props_appendix}.

Some intermediate-mass stars appear on the HRD diagram with ages younger/older than the chosen age boundaries of $0.4$~Myr and $5$~Myr, respectively, because their $T_{eff}$ and $L_{bol}$ estimates were obtained from the Virtual Observatory SED analyzer rather than from the $J$ versus $J-H$ color-magnitude diagram (see details in \S \ref{sec:stellar_props_appendix}).

The young stellar objects are separated into five source strata associated with specific loci on the HRD diagram:

\begin{enumerate}

\item Fully convective low-mass stars ($\la 1$~M$_{\odot}$) on Hayashi tracks, mainly M- and K-type stars (green symbols; 15,730 stars for the full sample and 237 stars for the super/mega-flare sample)   
\item Fully convective solar-mass stars ($1 \la M \la 2.5$~M$_{\odot}$) on Hayashi tracks, mainly K- and G-type stars (blue symbols; 7,058 and 303 stars)
\item Intermediate-mass stars ($2 \la M \la 5$~M$_{\odot}$) on Henyey tracks likely developing radiative cores. These include G-, F-, A-, and some late B-type stars (red symbols; 2,919 and 165 stars)
\item High-mass stars with $M \ga 5$~M$_{\odot}$, mainly B-type, stars (orange symbols; 974 and 44 stars).  Note that the X-ray emission probably is not produced by the massive primary, but rather by lower-mass unresolved secondaries in multiple systems as proposed for Orion Nebula Cluster B-type stars \citep{Stelzer05}.
\item Young stellar objects without available HRD locations and mass estimates (black symbols; 13,360 and 278).
\end{enumerate}

\begin{figure*}
\plotone{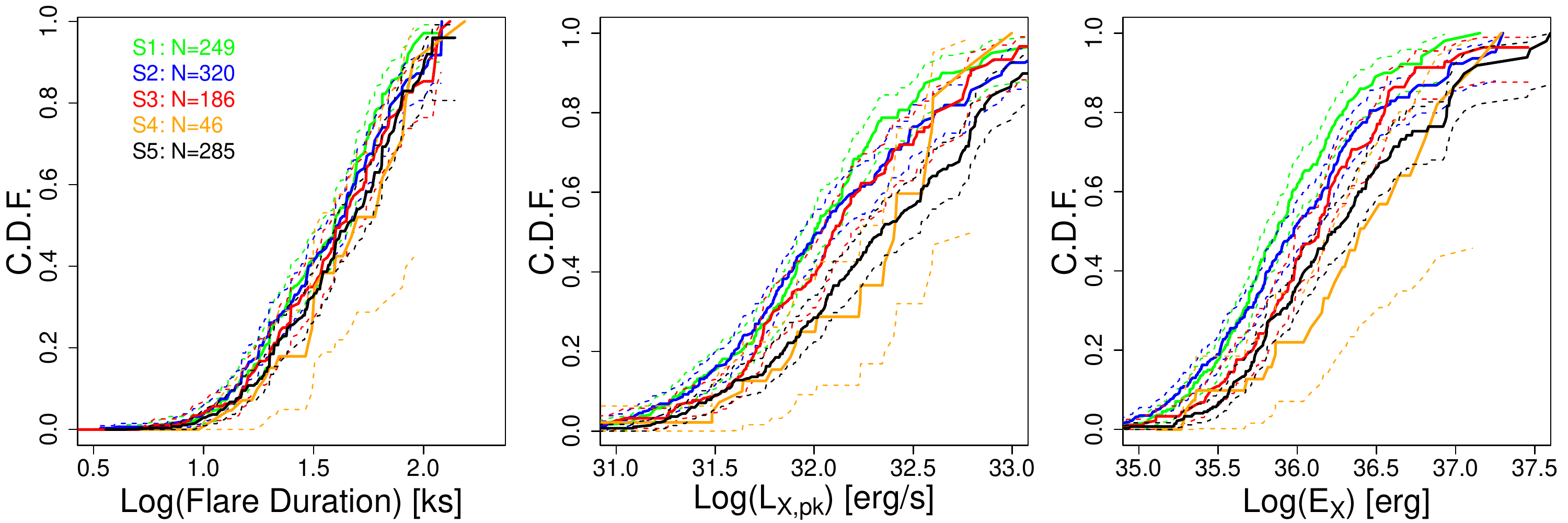}
\caption{Cumulative distribution functions  of the flare duration (left), X-ray flare peak luminosity (middle), and flare energy (right) constructed using Kaplan-Meier survival estimators. Dashed curves represent 95\% confidence bands for the KM estimators. Curve colors correspond to the five source strata from Figure~\ref{fig:super-flares_vs_allmystixsfincs}. The legends list the numbers of flares for each of the source strata. \label{fig:flare_energy_vs_mass}
}
\end{figure*}

Anderson-Darling nonparametric 2-sample tests between strata in each panel show significant differences for all of the full MYStIX+SFiNCs samples. For the super-flare hosts, low- and solar-mass strata (green and blue curves) have significantly different mass, radius, absorption and (with p-value $\sim$2\%) X-ray luminosity. Solar- and intermediate-mass strata (blue and red curves) differ in mass, radius and X-ray luminosity, but not absorption and median energy.  Intermediate- and high-mass strata (red and orange curves) differ in mass, (with p-value $\sim$5\%) radius, absorption and (with p-value $\sim$1\%) X-ray luminosity, but not median energy.

One finding from Figure~\ref{fig:super-flares_vs_allmystixsfincs} confirms a well-established result.  The X-ray luminosity pattern for super/mega-flare hosts is qualitatively similar to that of all MYStIX+SFiNCs stars, but with $L_{X,XPHOT}$ shifted towards higher values due to our selection of most powerful X-ray emitters. This strong correlation between the X-ray luminosity and stellar mass has been reported for young stars in many star forming regions \citep[e.g.,][]{Preibisch05,Telleschi07}.

Another confirmatory result is that young stellar objects across a wide mass range, from $0.1$~M$_{\odot}$ for M-type stars to $>5-10$~M$_{\odot}$ for B-type and even some O-type stars, produce X-ray super-flares. The X-ray flaring detected from massive stars may be associated with unresolved lower-mass stellar companions \citep{Stelzer05}. We find that $4-5$\% of solar-, intermediate-, and high-mass MYStIX+SFiNCs stars are in the super-flare sample. Only 1.5\% of the low-mass stratum produce super-flares, but this is expected from the  mass-$L_X$ correlation. The MYStIX/SFiNCs surveys have diminished sensitivities towards $<0.5-1$~M$_{\odot}$ stars and detect only a handful of young brown dwarfs. 

Two new results relating specifically to super/mega-flare host stars emerge from Figure~\ref{fig:super-flares_vs_allmystixsfincs}. First, the X-ray spectrum of super-flares from all mass strata appear indistinguishable.  This result emerges from the different X-ray median energy patterns between the full MYStIX+SFiNCs and the super-flare samples. For the full sample (as mentioned above), the lower-mass strata exhibit lower median energies than massive PMS stars \citep[a result known from Orion Nebula Cluster studies;][]{Preibisch05}; but the median energies are indistinguishable among the super-flare stars of different masses. This suggests that super/mega-flare physics and production mechanisms are similar across the wide range of star masses.

Second, the super-flare stars without mass estimates (black symbols) have much higher X-ray median energies indicating they are heavily absorbed. Half of these stars have median energies above 3~keV (equivalent to $\log N_H > 22.5$~cm$^{-2}$ or $A_V > 15$ mag) compared to only 15\% of stars with  identifiable mass estimates from the HRD. These young stellar objects have X-ray luminosity distributions similar to the visible $M>1$~M$_{\odot}$ samples. Most of them are associated with stellar clusters embedded in  molecular clouds \citep{Getman2018}. It is reasonable to infer that they are very young embedded objects, supporting the evidence in \S\ref{sec:protostars} that the production of X-ray super-flares starts very early in the PMS stages of evolution.

The X-ray properties shown in Figure~\ref{fig:super-flares_vs_allmystixsfincs} are based on the full {\it Chandra} exposure time, much of which may be associated with ``characteristic'' emission, composed of numerous weaker flares, before and after the super-flare.  In contrast, Figure \ref{fig:flare_energy_vs_mass} examines dependence of the super/mega-flare properties $-$ flare duration, peak X-ray luminosity, and total energy $-$ on host star mass. The figure shows Kaplan-Meier estimators of the cumulative distribution functions for the F+R+D super-flare subsamples stratified by stellar mass.  Note that the flare durations may be underestimated for flares lasting longer than $\sim 1$~day.  Several results are obtained from Figure~\ref{fig:flare_energy_vs_mass}.
\begin{enumerate}
\item Super/mega-flare duration distributions differ little from low-mass to intermediate- and high-mass stars. Nearly all lie between $\sim 20$~ks and $\sim 100$~ks with a median duration around 40~ks. This median duration for the MYStIX+SFiNCs super-flares is similar to that of the COUP super-flares \citep{Getman08a}.

\item Super/mega-flare $E_{X}$ distributions are correlated with stellar mass: the low-mass stratum is weaker than the intermediate-mass stratum with p-value $\sim 0.0003$ from the logrank test for equality of survival.  A similar effect in the $L_{X,peak}$ distribution may be present but is not statistically significant in our samples.  A reasonable explanation for a $E_X - M$ relation is that flare energy scales with the volume and footprint area of flaring loops involved in a single event, which in turn may depend on stellar surface area, hence on radius and mass. Another possibility is that the flare energy is powered by the strength of surface magnetic fields that may be stronger on stars with larger stellar volumes allowing more opportunity for a convective dynamo. A similar time-averaged $L_X - M$ relation, with similar possible explanations, is well-known in the Orion Nebula Cluster and other PMS populations \citep{Preibisch05}. 

\item The super/mega-flare duration, peak X-ray luminosity, and total X-ray energy distributions for the high-mass stratum (orange curve) consistently correspond to the longest and most powerful flares, albeit not all effects are statistically significant.  It is unclear why X-ray flares from B-type stars should be distinct from lower mass stars if the multiple stellar companion hypothesis for B-type star X-ray emission is correct \citep{Stelzer05}.

\item There are 450 MYStIX/SFiNCs `F+R+D' mega-flares above the completeness limit of $\log(E_{X}) = 36.2$~erg (\S \ref{sec:energy_distributions}). The ratios of mega-flares to mega+super-flares are 17\%, 26\%, 19\%, 6\%, and 32\% for S1, S2, S3, S4, and S5 sub-samples, respectively. These values are employed in the calculations of flare frequencies below.
\end{enumerate}

\section{Super- and mega-flare Occurrence Rates} \label{sec:flare_frequency}

\subsection{Measured and Extrapolated Mega-flare Occurrence Rates}

The frequency of super-flares in PMS populations, in units of flares per star per year, can be estimated from our analysis of the {\it Chandra} MYStIX+SFiNCs PMS sample as the ratio
\begin{equation}
    f_{SupFl} \simeq N_{SupFl} / (N_{PMS} * Med(t_{obs}))
    \label{eqn:flare_freq}
\end{equation} 
where $N_{SupFl}$ is the number of super-flares above a specified flare energy limit; $N_{PMS}$ is the total intrinsic pre-main sequence population observed with {\it Chandra}; and $Med(t_{obs})$ is the median of the total {\it Chandra} exposures among the observed 24,306 X-ray young stellar objects across the 40 MYStIX/SFiNCs star forming regions. However, careful estimation of these quantities is needed:

%\begin{enumerate}
    1. We treat $N_{SupFl}$ for incompleteness by considering here only the mega-flares where our sample is complete  (Figure~\ref{fig:cdf_kaplan_meier_full}). 
    
    2. $N_{PMS}$ is treated for  incompleteness in the {\it Chandra} MYStIX and SFiNCs PMS samples using the method of \citet{Kuhn15a} where the X-ray luminosity function (XLF) of each region is scaled to the Orion Nebula Cluster (\S~\ref{sec:datasets}, Table~\ref{tab:mystix_sfincs_regions}). $N_{PMS} \simeq 98,000$ and $14,000$ for the S1 and S2+S3+S4+S5 sub-samples, respectively. 
    
    3. The median {\it Chandra} observation exposure time among all the observed X-ray MYStIX and SFiNCs young stellar objects ($N=24,306$) across the 40 regions is $t_{Chandra} = 74.4$~ks with the bootstrap-derived 95\% confidence band of $\pm 0.1$~ks \citep{Kuhn2013a,Townsley2014, Broos13,Getman17}. 
%\end{enumerate}

\begin{figure}
\epsscale{1.2}
\plotone{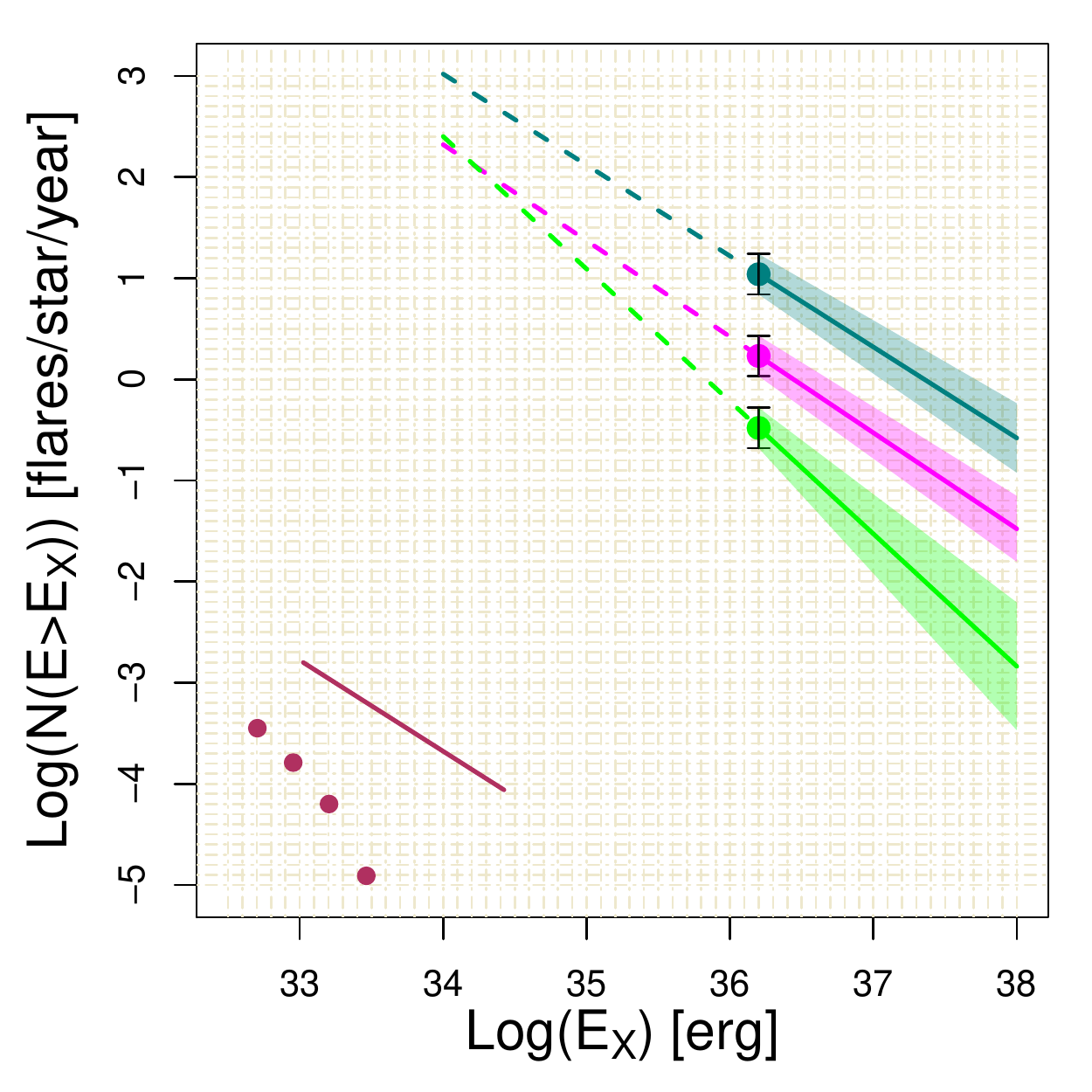}
\caption{PMS mega-flare occurrence rate as a function of flare energy. The measured MYStIX/SFiNCs mega-flare frequencies are the three points with $\pm 0.2$~dex error bars, representing three stellar mass ranges, in teal, magenta, and green.  The colored bands from $\log(E_X)=36.2 - 38$~erg represent the observed powerlaw mega-flare distribution, and the dashed lines extrapolate these relations to lower energies. Frequencies of optical band super-flares from older solar-type stars detected in {\it Kepler} satellite observations are shown in maroon, far below the PMS flare rates.  Further details are given in the text. \label{fig:flare_frequency}}
\end{figure}

Figure \ref{fig:flare_frequency} presents the result from equation \ref{eqn:flare_freq}: observed and extrapolated occurrence rates for PMS stars based on the MYStIX+SFiNCs mega-flares.  Teal and green indicate S2+S3+S4+S5 (more massive stars with $M>1$~M$_{\odot}$) and S1 (less massive stars with $M<1$~M$_{\odot}$), respectively, as discussed in \S\S \ref{sec:allmystixsfincs_vs_super-flares} and \ref{sec:flare_vs_mass}. Magenta indicates MYStIX/SFiNCs mega-flares averaged across the entire stellar mass range of $0.1-150$~M$_{\odot}$. 

The teal/magenta/green points with black error bars and the solid teal/magenta/green lines at high energies correspond to our direct measurements for super-flares exceeding energy completeness limit of $\log(E_X) = 36.2$~erg. The numbers of detected X-ray super-flares in these groups with energies above the completeness limit of $\log(E_X) = 36.2$~erg are $N_{SupFl} = 372, 450, 78$ flares, respectively (Figure~\ref{fig:cdf_kaplan_meier_full}). The error bars mark $\pm 0.2$~dex systematic errors on $\log(N_{PMS})$, as the largest error contributor in equation \ref{eqn:flare_freq}\footnote{For each of the three flare groups, the statistical errors on $N_{SupFl}$ and $Med(t_{obs})$ contribute less than $11$\% to the uncertainty on $f_{SupFl}$. The systematic error on $N_{PMS}$ due to the uncertainty in the methods used to derive total stellar populations, such as XLF versus initial mass function (IMF), is about 0.2~dex on $\log(N_{PMS})$ \citep[Figure 4 in][]{Kuhn15a}. This provides the largest contribution to the error on the inferred occurrence rate.}. 

Application of the methods from \S \ref{sec:energy_distributions} to the S1-green and S2+S3+S4+S5-teal samples yields Pareto slopes of $\beta = 1.31 \pm 0.24$ and $\beta = 0.90 \pm 0.08$, respectively.
The teal/magenta/green lines are the powerlaw relations with uncertainty envelopes bounded by the lines with the minimum (upper line) and maximum (lower line) powerlaw slopes; i.e., $\beta - err_{\beta}$ and $\beta + err_{\beta}$.  The color-coded lines in Figure~\ref{fig:flare_frequency} are:
\begin{equation}
\log N(E>E_X) = \kappa - \beta \times \log E_X,
\label{eqn:flare_freq_line}
\end{equation}
where $\kappa = $ 46.94, 34.62, 33.62 flares~(star-yr)$^{-1}$and $\beta = $ 1.31, 0.95, 0.90 for the low-mass (green), full IMF (magenta), and intermediate-mass (teal) flare groups, respectively.  The dashed lines extrapolate the frequencies with corresponding $\beta$ slopes to lower flare energies assuming no change in the powerlaw relations.

Frequencies of {\it Kepler} flares for solar-type stars are in maroon: the solid line is from \citet{Shibayama2013} and the four points are from \citet{Okamoto2020}.  Okamoto et al. revised the solar-type flare rates of the earlier study to lower values by accounting for the contamination from subgiants in their {\it Kepler} sample of flaring G-type main sequence stars. The optical flare energies from these studies were multiplied by $\times 1/15$ to give equivalent X-ray energies extrapolating the PMS optical-X-ray flare relation measured by \citet{Flaccomio2018} from simultaneous multi-band observations of a nearby PMS population.

The resulting frequencies of the MYStIX+SFiNCs mega-flares with $\log(E_X) > 36.2$~erg using equation (\ref{eqn:flare_freq}) are $0.3_{-0.1}^{+0.2}$, $ 1.7_{-0.6}^{+1.0}$, and  $11.0_{-4.1}^{+6.4}$~flares per star per year for the low-mass (green), full IMF (magenta), and high-mass (teal) groups, respectively. The mega-flare occurrence rates, at a fixed $E_X$ value, decrease with decreasing flare-host mass.

The MYStIX/SFiNCs sample includes roughly 6.5 times more $\log(E_X) > 36.2$~erg flares than the COUP flare sample of \citet{Colombo07}. Despite this large flare number difference, after re-normalization of the \citet{Colombo07} frequency from the total detected COUP point sources (1616, including contaminating AGNs) to the total stellar population in the COUP field of the Orion Nebula region ($N_{PMS}=1700$ stars; Table~\ref{tab:mystix_sfincs_regions}), their COUP flare frequency of $\sim 1.4$~flares per star per year is consistent with that of the full IMF MYStIX/SFiNCs mega-flare occurrence rate. 

The PMS flare occurrence rates inferred from Figure \ref{fig:flare_frequency} are remarkable!  The typical solar-mass PMS star is producing $\sim 1-3$ mega-flares~yr$^{-1}$ or $\sim 10^7$ mega-flares over the early $\sim 5$~Myr duration of the PMS evolutionary phase.  If the extrapolation of PMS mega-flares to lower unobserved energies around $\log(E_X) \simeq 34$~erg is valid, then PMS stars produce $\sim 10^6$ or more super-flares than older main sequence stars. Taking a typical flare duration of 50~ks (Figure~\ref{fig:flare_energy_vs_mass}), and assuming super-flares occur randomly in an ensemble of stars, a typical PMS star is experiencing a super-flare $\sim 30$\% of the time.

\begin{figure}
\epsscale{1.2}
\plotone{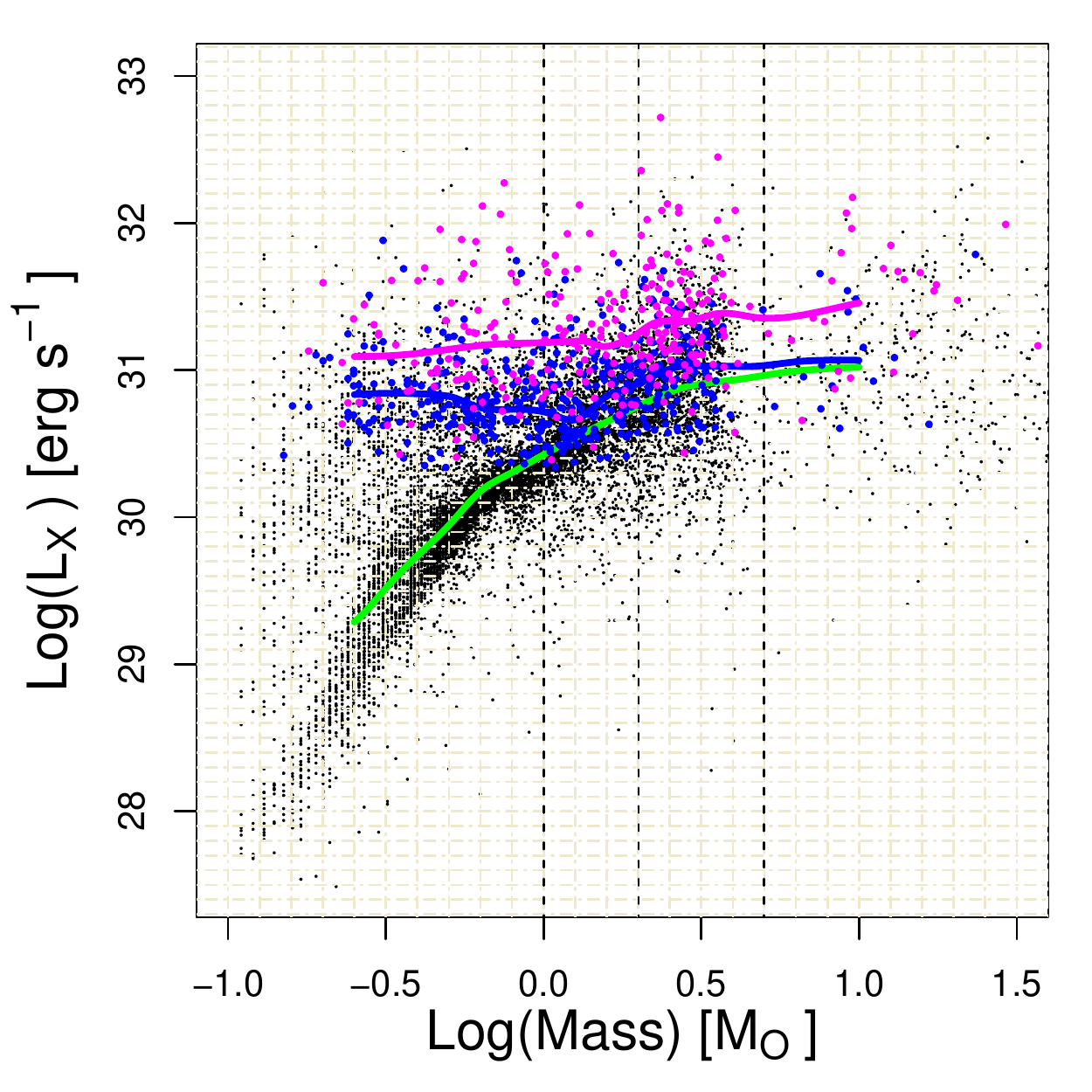}
\caption{X-ray luminosity of MYStIX/SFiNCs young stars as a function of stellar mass (black points). The super-flare hosts and their corresponding local regression fits are shown as points and curves in blue, and mega-flare hosts are shown in magenta. A local regression fit for the MYStIX/SFiNCs stars excluding the super/mega-flare hosts is marked by the green curve; these stars exhibit only ``characteristic'' X-ray emission. The dashed lines roughly delineate the mass ranges of the four source mass-strata described in \S \ref{sec:allmystixsfincs_vs_super-flares}. \label{fig:megaflare_contribution}}
\end{figure}

\subsection{Contribution of Mega-flares to PMS X-ray Fluence} 

A debate has waged for decades over the relative importance of powerful flares and nano- or micro-flares for heating the solar and stellar coronal \citep[][and references therein]{Nhalil2020}. While our study gives no information on the contribution of small flares, we can address the energy contribution of the mega-flares to the total time-integrated X-ray emission of PMS stars. 

One can assume that many weaker flares on a given star blur together into a quasi-continuous emission of X-rays that can be called the ``characteristic'' level  \citep[][]{Wolk05,Colombo07,Caramazza07}. For the MYStIX/SFiNCs young stars (excluding super-flare hosts), this characteristic emission is shown as a local regression fit (green curve) in Figure~\ref{fig:megaflare_contribution}. In the mass range $M>0.3$~M$_{\odot}$, this fit is similar to the $L_X - M$ relations for young stellar members of the Orion Nebula and Taurus regions reported by \citet{Preibisch05} and \citet{Telleschi07}\footnote{
Note they use earlier PMS stellar evolution models giving different mass estimates, especially for the lowest-mass range of $M<0.3$~M$_{\odot}$.}.

Figure \ref{fig:megaflare_contribution} shows that, for masses exceeding $\sim 1$~M$_{\odot}$ stars, super-flaring stars have levels of characteristic emission that are not unusually high compared to other PMS stars.  But at lower masses, super-flaring stars have unusually high time-integrated X-ray luminosities.  In contrast, mega-flaring stars are always have time-integrated X-ray luminosities far above of the characteristic emission of typical PMS stars in any mass range.

We concentrate here on the contribution of these most powerful mega-flares to the total X-ray fluence of PMS stars.    We integrate  mega-flare energetics for $36.2 < \log(E_X) < 38.0$~erg based on the flare frequency from Figure~\ref{fig:flare_frequency}, and we estimate the characteristic emission energetics based on the X-ray luminosities (green line) given in Figure~\ref{fig:megaflare_contribution}. Specifically, the mega-flare energetics (in ergs) released per year can be calculated as:

\begin{equation}
E_{tot} = \frac{10^{\kappa_{2}} \cdot \beta_{2}}{1-\beta_{2}} \cdot (E_{max}^{(1-\beta_{2})} - E_{min}^{(1-\beta_{2})}),
\label{eqn:tot_energy_per_year}
\end{equation}

\noindent where $E_{min} = 10^{36.2}$~erg and $E_{max} = 10^{38}$~erg (maximum energy of detected PMS flares). The $\kappa_{2}$ and $\beta_{2}$ parameters indicate the normalization and slope of the lower and upper boundaries in the uncertainty loci, shown as colored polygons in Figure \ref{fig:flare_frequency}. $\kappa_{2} = (55.43,38.45,36.95,32.29,36.32,30.94)$ and $\beta_{2} = (1.55,1.07,1.02,0.88,0.98,0.82)$ for the low-mass (green), full (magenta), and intermediate-mass (teal) flare groups, respectively.

Considering the median stellar masses and characteristic X-ray luminosities of flare host-stars for the three mass strata in Figures~\ref{fig:flare_frequency} and \ref{fig:megaflare_contribution}, the total mega-flare energies released per year are ($[0.8-3.2] \times 10^{36}$, $[0.7-2.0] \times 10^{37}$, $[0.5-1.4] \times 10^{38}$)~erg, respectively, compared to the total ``characteristic'' energies of ($2.7 \times 10^{37}$, $8.3 \times 10^{37}$, $2.2 \times 10^{38}$)~erg, respectively. Mega-flares with energies $\log(E_X)=[36.2-38]$~erg thus contribute about $3-11$\%, $8-19$\%, and $17-39$\% to the total X-ray energetics of the $\la 5$~Myr old PMS stars, in the low-mass (green), full IMF (magenta), and intermediate-mass (teal) flare groups, respectively. As with the occurrence rate, the contribution of mega-flares to the total PMS energetics decreases with decreasing flare-host mass.

Extrapolation of the PMS flare occurrence rate (equation \ref{eqn:flare_freq_line}) towards higher energies, allows statistically to have monster PMS flares with energies up to $\log(E_X) = 10^{42}$~erg over the first few Myr of PMS evolution. However, the PMS magnetic fields \citep{Sokal2020} may not be sufficiently strong to provide such enormous flaring energy. For solar-type flares, up to 10\% of magnetic energy can be converted to flare energy \citep{Okamoto2020}. One can imagine a PMS star with a radius $R_{\star}=2.5$~R$_{\odot}$ hosting a large active region, which covers a quarter of the stellar surface. The region is powered by surface magnetic field of $B=5000$~G (currently the maximum measured average PMS field \citep{Sokal2020}) with magnetic energy stored in a volume reaching depth of $0.1$~R$_{\star}$. The total magnetic energy would be $\sim B^2 \cdot V /8\pi \sim 2 \times 10^{39}$~erg, insufficient to power monster flares. Extreme surface magnetic fields reaching up to $20$~kG for cases of powerful flares from most magnetically active stars are predicted by recent theoretical calculations \citep{Zhuleku2021}. Such fields could provide total magnetic energy of $3 \cdot 10^{40}$~erg, perhaps allowing some monster flares. Since these ideas are only speculative and semi-quantitative, we restrict the analysis of the mega-flare energetics contribution to the observed value of $E_{max} = 10^{38}$~erg. These fractional contributions to the characteristic X-ray emission could be higher if flaring extends to energies above the  $\sim 10^{38}$~erg maximum observed here.

\section{Comparison of X-ray and Optical Band super-flares} \label{sec_optical}

\subsection{Main Sequence and PMS Stars \label{sub_sec_d1}}
 
\citet{Ilin2021} summarize recent flare surveys using {\it Kepler}, {\it TESS}, Evryscope, and other optical photometry \citep{Lurie2015,Chang2015,Ilin2019,Lin2019,Raetz2020,Davenport2020}. The power-law slopes of these flare distributions $dN/dE \propto E^{-\alpha}$ are consistent with our slope of $\alpha = 1.95 \pm 0.07$ (\S \ref{sec:energy_distributions}). However, flare energies from most of these surveys do not exceed $\log(E_{opt,f}) < 34-35$~erg. Only the {\it Kepler}-detected flares from G-type dwarfs \citep{Shibayama2013} reach energies up to $\log(E_{opt,f}) \sim 36$~erg. \citet{Notsu19} show that flares from relatively younger ($t < 500$~Myr) G-type stars are more likely to reach energies of up to $10^{36}$~erg.  

The bulk of solar flare energy is generally associated with the optical continuum rather than the soft X-ray band \citep{Schrijver2012}. For young star flares with $\log E < 36$~erg, the flare energy in the optical continuum exceeds the energy in the soft X-ray band on average by a factor of $\sim 6$ \citep{Flaccomio2018}.  

Thus none of these optical studies targeting main sequence stars have flares with energies comparable to our X-ray mega-flares in the energy range $\log(E_{X,f}) \sim 36-38$~erg, and most of the observed main sequence flares are weaker than our X-ray super-flares.  This comparison agrees with \citet[][Figures 8-9]{Getman08a} indicating that the COUP super-flares are the most powerful among known solar-stellar flares \citep{Gudel2004, Aschwanden2008}. 

For PMS stars, \citet{Jackman2019} report detection of a single white-light mega-flare from the nearby, 2~Myr old, M-type star NGTS J121939.5-355557 with flare energy of $\log(E_{opt,f}) \sim 36.5$~erg. They estimate a crude flare occurrence rate of $0.2-3$ flares~yr$^{-1}$ for flares with $\log(E_{opt,f}) > 36$~erg.  Assuming the factor of 6 optical-to-X-ray energy ratio from \citet{Flaccomio2018}, the crude estimate of the flare frequency for the MYStIX/SFiNCs low-mass stars with X-ray energies of $\log(E_{X,f}) > 35.2$~erg (green dashed line in Figure \ref{fig:flare_frequency}) is of several flares per star per year, similar to, but somewhat above, the occurrence rate estimated by Jackman et al.

Note that it is possible that optical super-flares would appear less frequent than X-ray super-flares for a star with a given flaring rate due to loop geometry.  The optical emission is probably emitted at loop footprints near the photosphere whereas X-ray emission is probably emitted throughout the loop that is much larger than a stellar radius \citep{Getman08b}.  As only one hemisphere of the photosphere is visible at a given time,  the apparent occurrence rate of optical super-flares could be $\sim 2$ times lower than the occurrence rate of X-ray super-flares. More discussion of this issue can be found in \citet{Flaccomio2018} who detected several dozen bright X-ray flares with optical and/or mid-IR flare counterparts in PMS members of the NGC~2264 star forming region. Their estimated fraction of X-ray flares with no optical counterparts varies with the choice of flare morphology, brightness, and energetics from 19\% to 48\%.   

\subsection{Super-flare Rate Dependence on Stellar Age \label{sub_sec_d2}}

In light of extrapolations of flare energies and occurrence rates over wide ranges, and for different flare samples at different bands with different sensitivities, the flare rates as functions of age estimated below can only be considered suggestions rather than reliable results.

The {\it Kepler} and {\it TESS} monitoring of the young ($30-50$~Myr) M-type star GJ~1243 \citep{Davenport2020} gives an occurrence rate of $E_{opt,f}>10^{34}$~erg flares as roughly 10~flares (star-yr)$^{-1}$. For this flare energy range, the extrapolated optical-to-X-ray energy ratio is around 20 \citep{Flaccomio2018}, giving an X-ray energy of $\log(E_X) \sim 32.7$~erg. Extrapolation of the MYStIX/SFiNCs trend for low-mass PMS stars with ages $<5$~Myr (green line in Figure~\ref{fig:flare_frequency}) down to $\log(E_{X,f}) \sim 32.7$~erg gives a rate of $\sim 13,000$~flares per star per year. Comparing to the rate obtained for GJ~1243, the super-flare frequency rate may decrease by a factor of  $\sim 1300$ between the ages of $t \la 5$~Myr and $30-50$~Myr. However, this rough estimate is based on a single star studied in the optical band in contrast to the large sample we have in the X-ray band for PMS stars.   

Based on the optical flares collected by \citet{Ilin2021}, the flare rate for $\sim 135$~Myr old low-mass stars in the Pleiades cluster at $E_{opt,f} > 10^{34}$~erg is $\sim 0.4$~flares (star-yr)$^{-1}$. This suggests the super-flare frequency from young low-mass stars drops by a factor of $\sim 32,000$ between the ages of $t \la 5$~Myr and $t \ga 100$~Myr. This is consistent with the observed X-ray super-flare energetics decrease between the Orion and Pleiades populations \citep{Guarcello2019}. Our current large {\it Chandra} survey of a dozen $10-100$~Myr old stellar clusters employing homogeneous datasets and methods will further help to quantify flare frequencies at this stellar evolutionary period (Getman et al., in prep.).   

The tentative results for MYStIX/SFiNCs low-mass PMS stars presented here suggest that super- and mega-flare occurrence rates with $\log(E_{X,f}) \gtrsim 32.7$ erg declines steeply during the early phases of stellar evolution by a factor of $\sim 1000$ from 5 to 50 Myr and a factor of $\sim 30,000$ from 5 to 135 Myr. Thus, the occurrence rate trend is roughly $t^{-3}$ over this age range. \citet{Ilin2021} report that optical flare rates in open clusters continue to decline over the age range $\sim 100 - 3000$~Myr. A recent estimate of the super-flare occurrence rate on the Sun is $\sim 0.0002$ flares~yr$^{-1}$ at a level of $\log(E_{opt,f}) = 34$~erg ($\sim$X1000-class) \citep{Okamoto2020}.

\section{Effects of PMS super-flares on the Environs \label{sub_sec_d3}}

\subsection{Implications for Protoplanetary Disk Photoevaporation}

There is little doubt that PMS X-rays efficiently irradiate circumstellar disks during the early PMS phase. X-ray photoevaporation models reproduce reasonably well the line profiles of the [OI] 6300 line as a tracer of warm quasi-neutral disk wind \citep{Picogna2019}. \citet{Gudel10}  `` ... find indications that the production of [NeII] emission weakly scales with the X-ray luminosity'' supporting models of disk irradiation by X-rays. \citet{Flaischlen2021} report a negative correlation between the X-ray luminosity and accretion rate for similar mass/age ONC stars as a signature of the X-ray driven disk photoevaporation. Recurring powerful X-ray flares are also proposed to sustain the observed extended [NeIII] emission in the jets of young stars, such as DG Tau \citep{LiuChunFan2016}.

Disk photoevaporation linked to high-energy radiation from the central star is now directly detected in a number of systems, well-explained by hydrodynamical calculations of the photoevaporative winds driven by stellar ultraviolet and X-ray emission \citep{Alexander2014,Picogna2019}.  X-rays seem most important in the final stages of disk dispersal \citep{Owen13}. 

Since the MYStIX+SFiNCs cluster samples range in age from $<0.5$ to $\sim 5$~Myr \citep{Getman14a}, and since protoplanetary disks have longevities around 2-8~Myr \citep[][and references therein]{Richert18}, we can directly estimate the total number of super-flares that have irradiated a typical disk over a $\sim 5$~Myr lifetime from the occurrence rates shown in Figure~\ref{fig:flare_frequency}.  {\it The result is impressive: Disks around PMS stars with masses $\sim 1$~M$_\odot$ will be} irradiated by $\sim 1$ billion super- and mega-flares with energies $34 < \log(E_X) < 38$ erg.  

If the power law distribution extends to much higher energies, which may not be true, then a disk around a solar mass PMS star would experience $\sim 2 \times 10^5$ flares with energies $\log(E_X) > 38$ erg. Disks around low mass PMS stars that will become the populous dM dwarfs will be irradiated by $\sim 1 \times 10^9$ super/mega-flares with energies $34 < \log(E_X) < 38$ erg, and possibly by $\sim 7 \times 10^3$ flares with energies $\log(E_X) > 38$ erg.

Photoevaporation from X-ray super/mega-flares may affect planet formation processes.  In the early stages, flare X-ray ionization will briefly penetrate into the middle $-$ and possibly midplane $-$  disk layers.  The plasma producing X-rays in these powerful events is unusually hot compared to plasma produced in solar magnetic reconnection events with peak temperatures ranging from 20 to 100 MK and higher \citep[][and Getman, Feigelson, \& Garmire, 2021, ApJ, submitted]{Getman08a, Getman08b, Getman2011}.  The bremsstrahlung spectrum from protostellar flares less luminous than our super-flares has been detected out to  energies of $\sim 20$~keV \citep{Vievering19}. X-ray penetration into solar-abundance gas scales approximately with the cube of photon energy and can attain column densities $\log(N_H) \sim 25-26$ cm$^{-2}$ \citep{Glassgold2000}, sufficient to reach the mid-plane in the outer regions of some disks \citep{Ilgner06}.  Super-flare ionization thus has the potential to increase turbulence from magnetorotational instability and induce ion-molecular chemistry in disk interiors.  However, the importance of the effect depends critically on recombination rates that are difficult to estimate.  

In the later stages, the removal of gas from a protoplanetary disk by flare X-ray photoionization may be sufficient to trigger the streaming instability that rapidly forms pebbles and planetesimals critical to the rapid formation of protoplanets \citep{Lambrechts12, Carrera2017}.  For instance, the gas in disks with initial sizes of 30~AU and 100~AU around a $1$~M$_{\odot}$ young star can be removed within 1~Myr and 4~Myr, respectively \citep{LiuLambrechtsJohansen2019} based on a gas removal rate by the ``characteristic'' X-ray emission component $\dot{M}_{phot} = 6 \times 10^{-9} \times (L_X/10^{30})^{1.14}$ of $1.8 \times 10^{-8}$~M$_{\odot}$~yr$^{-1}$ \citep{Owen2012}. The mega-flare ($\log(E_{X,flare}) > 36.2$~erg) X-ray component alone (\S \ref{sec:flare_frequency}) would increase this rate by $>10-20$\% and consequentially speed up processes of planetesimal and planet formation.  Detailed astrophysical calculations are needed to more reliably estimate possible nonlinear effects of short-lived super-flare irradiation on disk.  

\subsection{Implications for Protoplanetary Disk Spallation and Chemistry\label{sec_other_effects}}

After the discovery of X-ray flaring in PMS stars, it was proposed that energetic particles ($E \ga 10$~MeV) associated with these magnetic reconnection events could have produced short-lived radionuclides in the solar nebula through nuclear spallation \citep{Feigelson82}.  Today the evidence indicates that, while many meteoritic radionuclides arose from supernova explosions near the Sun's natal molecular cloud, some radionuclides were formed by spallation from an `early active Sun' with the elevated flaring behavior seen in PMS stars \citep{Chaussidon06}. Excess $^{10}$Be in Ca-Al-rich inclusions (CAIs) is particularly important as sufficient quantities cannot form in supernova events \citep{McKeegan00}. $^{10}$Be abundances vary widely among CAIs and were produced after $^{26}$Al from supernovae decayed; these and other properties (e.g., covariation with $^{50}$V, irradiation products in $^{26}$Al-free hibonite-rich CAIs) point to a spallogenic origin by solar energetic particles rather than a pre-solar source distributed throughout the disk \citep{Fukuda2019}.  

Assuming ``characteristic'' X-ray emission around $\log(L_X) \sim 30$ erg~s$^{-1}$, a rough estimate is that PMS  proton fluence is elevated $\sim 10^5$ above contemporary solar levels \citep{Feigelson02}.  This would be sufficient to produce the observed abundances of spallogenic radionuclides  \citep{Rab17}.  The scaling of energetic proton flux to X-ray luminosities for super-flares is unknown, so quantitative estimates of the effects of super-flare particles on radionuclide production can not be made at this time. 

X-ray ionization should induce ion-molecular chemistry, and an unusual case of variable HCO$^+$ emission can be attributed to X-ray flaring. \citet{Cleeves2017} report variability in  H$^{13}$CO$^{+}$ line emission in the disk of PMS  IM Lup.  The implied rapid abundance changes of the HCO$^{+}$ molecular ion can be explained by X-ray flaring that ionize the H$_{2}$ gas on the disk surface.  This  produces H$_{3}^{+}$ followed by the proton transfer reaction with CO to produce HCO$^{+}$ ions. Their flare-driven disk chemistry simulations involving X-ray flares with energies $\sim 1 \times 10^{36}$~erg results in enhanced HCO$^{+}$ abundances for a period up to $\sim 20$~days.

Simulations of X-ray super-flare driven chemistry in a disk around a young solar-mass star also predict  changes in the gas-phase H$_{2}$O abundance lasting days \citep{Waggoner2019}. Their choice of super-flare frequency for $\log(E_X) = 37.1$~erg flares once ``every few years'' based on the COUP studies is consistent with our rate of 1~flare per star per 4 years (magenta line in Figure~\ref{fig:flare_frequency}).

Theoretical calculations predict many other effects of stellar X-ray irradiation on the disk, accretion and outflow astrophysics.  These include: stimulation of the magnetorotational instability and associated turbulence \citep{Fromang02}; ionization necessary for launching a magnetocentrifugal disk wind \citep{Gressel13} and a collimated jet \citep{Shang2002};  desorption of water ice from dust grains \citep{Dupuy18}. The magnetohydrodynamic simulations of \citet{Colombo2019} suggest that super-flares may trigger formation of accretion funnels and influence morphology of inner disk and accreting columns. 
\subsection{Implications for Young Planetary Atmospheres \label{sub_sec_d4}}

Evidence for early Jovian planet formation emerges from a number of recent observations \citep[][and references therein]{Liu2020} including: ALMA detections of compact rings and gaps in $\la 1$~Myr old disks \citep{Andrews2018}, optical- and IR-band radial velocity detections of hot Jupiters around the $\sim 2$~Myr old PMS stars CI Tau and V830 Tau \citep{Johns-Krull2016, Donati2016}, and direct $H_{\alpha}$ imaging of accreting proto-planets within the transition disk of the $\sim 5$~Myr old star PDS 70 \citep{Haffert2019}.

Recent theory based on the streaming instability and pebble accretion indicates that the formation of rocky super-Earths, necessary for the gravitational trapping of disk gas to form Jovian planets, may be extremely rapid \citep{Raymond20}.  This can be followed by migration of resonant chains to the inner edge of the disk where the nascent rocky planets can be subject to intense radiation from super-flares.  Most of the resonant chains become unstable when the disk dissipates but many compact planetary systems survive.  These inner rocky planets may have water-rich volatile atmospheres.  

Although the super/mega-flare occurrence rate rapidly decreases as the PMS star enters its main sequence phase (\S\ref{sub_sec_d2}), the flares continue after the disk dissipates and can no longer protect the inner planets from flare higher energy photon irradiation.  In addition, the planetary atmospheres can be impacted by coronal mass ejections with much greater total plasma energy than the radiative energy from super/mega-flares.  

\citet{Poppenhaeger2020} describe a relevant calculation for the four-planet system around the 20~Myr old solar-mass PMS star V1298 Tau.  The innermost planet {\it c} orbiting 0.08~AU (17~R$_\odot$) from the star may lose a hypothetical H/He envelope within $\sim$100~Myr. If such a planet was orbiting a young ($\la 5$~Myr) $1$~M$_{\odot}$ star with ``characteristic'' X-ray luminosity of $L_X = 2.5 \times 10^{30}$~erg~s$^{-1}$ (\S \ref{sec:flare_frequency}), then the hydrodynamic escape assumption of $\dot{M} = (\epsilon \pi R_{XUV}^2 R_{pl} F_{XUV})/(K G M_{pl})$ \citep{Owen2012} with the atmospheric escape efficiency $\epsilon = 0.1$, the Roche lobe factor of $K = 0.8$, the ``fluffy'' planetary radius at XUV wavelengths of $\times 2$ of radius at optical wavelengths ($R_{pl} = 5.6$~R$_{\earth}$), and the conservatively chosen EUV flux as $4 \times F_{X-ray}$ \citep{Sanz-Forcada2010}, would result in the H/He envelope mass loss rate of $0.11$~M$_{\earth}$~Myr$^{-1}$ and complete evaporation of the envelope within $4.5$~Myr.  

A similar calculation by \citet{Johnstone2019} suggests even faster destruction of early planetary atmospheres.  Considering the effects of extreme ultraviolet irradiating planets around a $\sim 100$~Myr solar-mass star, they find removal of even heavy-element atmosphere on timescales of $\sim 0.1$~Myr by photodissociation and/or hydrodynamic escape. 

The addition of the super/mega-flare X-ray emission component would further shorten this atmosphere evaporation process. If the disk gas is removed after $\sim 5$~Myr, the young planets will experience roughly one billion flares with energies $34 < \log(E_X) < 38$~erg, including several million mega-flares with energies $32.6 < \log(E_X) < 38$~erg.  The mega-flare effects may be modest ($10-20$\%; \S \ref{sec:flare_frequency}) if an intense burst of X-rays has the same effect as a continuous irradiation of characteristic X-ray emission. But short-lived super-flares may have nonlinear effects. Even at older ages, \citet{Atri20} find that the super-flare emission may dominate envelope loss for $\sim 20$\% of late M-type stars.  New calculations are needed to evaluate whether short-lived intense bursts of X-rays have the same evaporative effects as a weaker but continuous irradiation of X-rays. 

Super-flares may have other effects on young planetary atmospheres.  Their ozone layer may be depleted or destroyed by stellar energetic particles leading to increased penetration of ultraviolet radiation to the planetary surface \citep{Schaefer00, Tilley2019, Howard2019}.  Super-flare energetic particles may stimulate non-equilibrium atmospheric chemistry such as the production of nitrous oxide and hydrogen cyanide \citep{Airpetian2016}. This conceivably might promote surface organic chemistry leading to the formation of life. Finally, energetic super-flare photons may improve, rather than destroy, the effectiveness of photosynthesis in inhabited zone planets around late-M stars \citep{Mullan2018}.

\section{Concluding Remarks} \label{sec_concluding_remarks}

The $Chandra$ MYStIX and SFiNCs surveys have produced a sample of   $>$30,000 X-ray emitting PMS stars with ages $<5$~Myr from 42 star forming regions within $d < 3$~kpc in the Galactic disk. Omitting the Carina Nebula and Orion Nebula regions, here we examine the X-ray variability among the remaining $>24,000$ X-ray young stars.  Using a reproducible likelihood-based statistical procedure (\S\ref{sec:methods} and Appendix~\ref{sec:change_point_model}), we extract over a thousand flares.  An atlas of the flare lightcurves and properties is provided (\S\ref{sec:flare_atlas}).  Peak luminosities lie in the range  $\log(L_X)=30.5-34.0$~erg~s$^{-1}$ with total energies $\log(E_X)=34-38$~erg in the $Chandra$ $0.5-8$~keV band.  The sample is `complete' above $\log(L_X) > 32.5$ erg~s$^{-1}$ and $\log(E_X) > 36.2$ erg (\S\ref{sec:energy_distributions}).  This is the largest collection of powerful stellar flares ever assembled in the X-ray band.  They are far more luminous than the optical band flares recently studied in main sequence stars (\S\ref{sec_optical}).  

We highlight here two themes where the findings influence important astrophysical issues.  

\subsection{Super/mega-flares Do Not Arise from Star-Disk Magnetic Fields} 
A basic result of our study is the ubiquitous nature of the X-ray super-flares among PMS stars.  Averaged over an IMF ensemble, each PMS star produces several super-flares (with energies $E_X>10^{34}$~erg) per week and 1-3 mega-flares ($E_X>10^{36.2}$~erg) per year (\S \ref{sec:flare_frequency}). Since the former flare frequency number is inferred using extrapolation towards lower energies, beyond our completeness limit (dashed magenta line in Figure \ref{fig:flare_frequency}), it should be considered with caution. 

Super/mega-flares occur in all star forming regions, from stars of all masses and at all stages of young stellar evolution.  The flares are seen in heavily absorbed Class I protostars with enormous infrared excesses from large protoplanetary disks (\S~\ref{sec:protostars}), in Class II T Tauri stars that are still accreting from their disks, and in Class III diskless stars.  The collective super-flare energetics are indistinguishable across evolutionary classes (\S \ref{sec:disky_diskless}). Our companion paper will show that super-flare astrophysical properties (such as peak luminosities, decay timescales, and plasma temperatures, densities and volumes) similarly are indistinguishable between disk-bearing and diskless stars (Getman, Feigelson, \& Garmire, 2021, ApJ, submitted).  

Our analyses thus provide no evidence for a distinct flaring mechanism involving the circumstellar disk, such as reconnection in field lines connecting the star and disk, or at the boundary between the stellar magnetosphere and the inner disk.  Such mechanisms have been speculated from the pioneering scenarios of \citep{Hayashi1996} and \citet{Shu1997} to recent 3D+time magnetodynamical calculations of \citet{Colombo2019}.   The only remaining links between X-ray emission and disks are indirect, such as the possibilities that X-ray loop sizes are constrained to lie within the inner disk boundary \citep{Getman08b}, star-disk loops are responsible for a periodic variation in X-rays seen in two ONC super-flares \citep{Reale2018}, and X-ray flares trigger increased accretion from the inner disk \citep{Espaillat19}.   If there are magnetic reconnection events involving star-disk magnetic field lines, they do not manifest themselves in detectable super/mega-flares, and/or the occurrence rate of such star-disk events is very low.  

The similarity of coronal X-ray properties of disk-bearing and diskless stars has been seen in many studies from early observations with $ROSAT$ \citep{Feigelson93} to thorough studies with $XMM-Newton$ \citep{Gudel07} and $Chandra$ \citep{Preibisch05}. Studies of the Taurus \citep{Stelzer07}, NGC 2264 \citep{Flaccomio2018}, Orion Nebula \citep{Getman08b,Flaccomio2012}, and MYStIX/SFiNCs regions (here) provide observational evidence that young stars with and without disks produce flares with similar properties (\S \ref{sec:disky_diskless}). The proposed physical processes responsible for the mild suppression of time-integrated X-ray emission in accreting versus nonaccreting PMS systems, as well as the accretion shocks producing relatively weak soft X-ray excess emission (\S \ref{sec:intro}) seem to have little or no effects on the production mechanisms and characteristics of coronal flaring including super-flares of extraordinary power with $\log(L_X)=30.5-34.0$~erg~s$^{-1}$ and total energies $\log(E_X)=34-38$~erg. 

When combined with solar flare properties \citep[e.g.,][]{Aschwanden2008}, the observational evidence acquired in our own studies consistently points to solar-type geometries, magnetic loops with both footprints rooted in the stellar surface, analogs of giant solar X-ray arches and streamers \citep{Getman08a,Getman08b}.  The flare processes span a phenomenal range: 2-3 orders of magnitude in flare duration; 4 orders of magnitude in loop length; and 13 orders of magnitude in plasma emission measure \citep{Getman2011}.  Often the same distribution functions (\S\ref{sec:energy_distributions}) and scaling relations are seen for both weak and powerful flares. 

However, rare examples of star-disk flaring may have been found.  \citet{Reale2018} report detection of quasi-periodic pulsations in COUP super-flares produced by two disk-bearing young stars, V 772 Ori\footnote{In our opinion, V 772 Ori exhibits properties consistent with a diskless system: EW(CaII)$=1.4$, $\Delta(H-K_s)=-0.19$~mag, [3.6]$-$[4.5]$=0.06$~mag \citep{Getman05,Getman08a} and apparent {\it Spitzer}-IRAC SED slope $\alpha_{IRAC}=-2.5\pm0.1$ (based on the full {\it Spitzer} Orion point source catalog of Megeath; \url{http://astro1.panet.utoledo.edu/~megeath/megeath_group/The_Spitzer_Orion_Survey.html}).} and OW Ori. Their hydrodynamic flare modeling suggests single flaring loops with sizes significantly exceeding corotation radii, supporting  star-disk loop configurations. While individual star-disk flares may exist, the majority of PMS super/mega-flare events appear to be associated with magnetic loops anchored in the stellar surface. 

\subsection{Super/mega-flares Contributions to Disk and Protoplanet Irradiation}

Although protoplanetary disks and planets are neutral molecular phases with typical thermodynamic temperatures in the range $100-2000$~K, it is widely recognized that they are impacted by external high energy radiation with potentially enormous effect. PMS stellar X-rays constitute only $10^{-3}$ of their bolometric luminosity and therefore have little effect on heating of the bulk circumstellar material.  But their ionization induce disk turbulence and non-equilibrium ion-molecular chemistry, trap mostly-neutral material to magnetic field lines for accretion or outflows, and photoevaporate outer layers of disk or planetary atmospheres.  Furthermore,  undetected bursts of energetic particles and coronal mass ejections are likely to accompany super/mega-flares with additional effects on circumstellar gaseous and solid material. 

However, it is difficult to quantify these effects due to the uncertain timescales of the astrophysical response to the flare event.  In \S\ref{sec:flare_frequency}, we find that the time-integrated ensemble X-ray luminosities of PMS stars is elevated at least $10-20$\% by mega-flares.  This arises from the steep $\alpha \simeq 2$ slope in the flare energy distribution (\S\ref{sec:energy_distributions}). If the response of disk ionization, planetary atmosphere escape, or other effect is much slower than the super-flare timescale, then the total effect of the super-flares will be modest.  

But if the response is rapid, then the effects can be substantial. The observation of variable HCO$+$ emission in a disk by \citet{Cleeves2017}, and the calculation of rapidly fluctuating disk `active' and `dead' zones by \citet{Ilgner06}, suggest that ionization effects can be sufficiently rapid  that the astrophysical response is strong.  More astrophysical modeling of the time dependency of such effects is needed to gain confidence in any conclusion concerning the importance of super/mega-flares on disk and planetary processes.  

\acknowledgments
We are grateful to the referee for spending her/his time and providing many useful suggestions that stimulated fresh ideas and improved the paper. This project is supported by the {\it Chandra} archive grant AR9-20002X and the {\it Chandra} ACIS Team contract SV474018 (G. Garmire \& L. Townsley, Principal Investigators), issued by the {\it Chandra} X-ray Center, which is operated by the Smithsonian Astrophysical Observatory for and on behalf of NASA under contract NAS8-03060. The {\it Chandra} Guaranteed Time Observations (GTO) data used here were selected by the ACIS Instrument Principal Investigator, Gordon P. Garmire, of the Huntingdon Institute for X-ray Astronomy, LLC, which is under contract to the Smithsonian Astrophysical Observatory; Contract SV2-82024. This research has made use of NASA's Astrophysics Data System Bibliographic Services.

%\vspace{5mm}
\facilities{CXO}

%\software{astropy \citep{2013A&A...558A..33A},  
%          Cloudy \citep{2013RMxAA..49..137F}, 
%          SExtractor \citep{1996A&AS..117..393B}
%          }

% Appendix
%\newpage
%\vspace{100mm}
%%%%%%\lineskip
\appendix
\newpage

\section{Energetics Of Brightest COUP Flares} \label{sec:coup_flare_energies}

\citet{Getman08a,Getman08b} examined and calculated  the properties of flaring coronal structures associated with the sample of the brightest 216 flares from COUP young stars. This is the largest tabulated sample of PMS super-flares with modeled loop geometries. In our companion paper (Getman, Feigelson, \& Garmire, 2021, ApJ, submitted), we model the brightest MYStIX/SFiNCs super-flares and compare their inferred properties with those of the COUP flares from \citet{Getman08a,Getman08b}. 

To better understand the energetics of these COUP flares, here we use their reported peak flare X-ray luminosity ($L_{X,pk}$) and decay $e$-folding timescale ($\tau_d$) of flare X-ray counts \citep[their Table~2;][]{Getman08a} to estimate flare energies $E_X \simeq L_{X,pk} \times \tau_d$.  We fit the unbinned energies with the Pareto (powerlaw) function using the standard maximum likelihood procedure \cite[e.g.,][]{Newman05}. Most COUP flares have full observation coverage thanks to the extremely long COUP exposure time; hence no need for a Kaplan-Meier estimator.

\begin{figure*}[ht]
\plotone{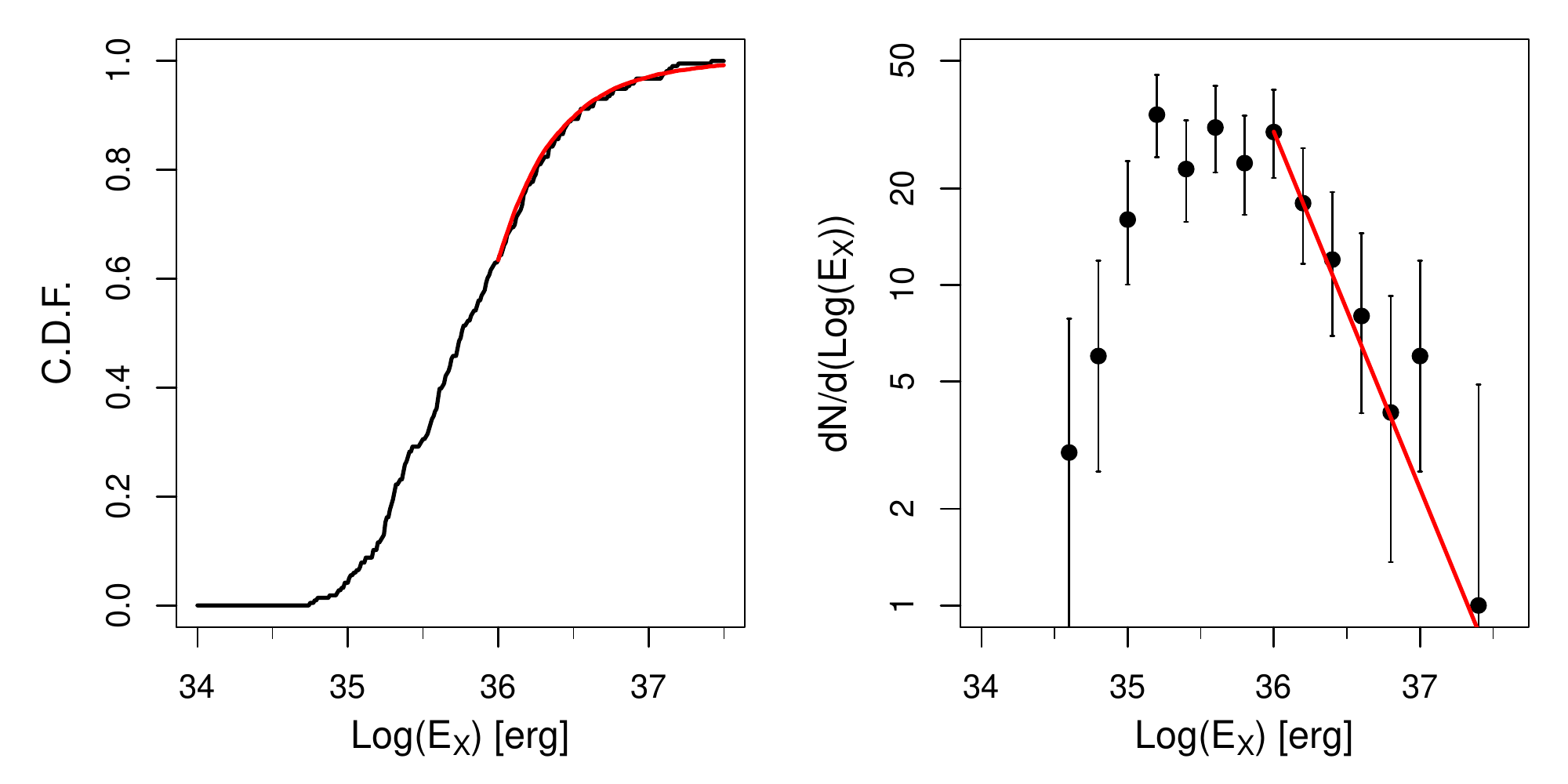}
\caption{Cumulative distribution function of flare energies ($E_X$, left) and corresponding histogram visualization of the probability distribution function with 0.2 dex bins (right) for 216 bright flares in the COUP observation of the Orion Nebula Cluster (data from \citet{Getman08a}). The best-fit Pareto (powerlaw) function fit to the high end of the c.d.f. is shown in red.  Histogram error bars are approximations to 95\% confidence intervals of a Poissonian distribution \citep{Gehrels86}.\label{fig:cdf_coup}}
\end{figure*}

Figure~\ref{fig:cdf_coup} shows that this 216 COUP flare sample is complete above $\log(E_X) \geq 10^{36}$~erg with the powerlaw shape of $dN/dE_{X} \propto E_X^{-2.1}$, or equivalently $dN/d \log E_X \propto E_X^{-1.1}$. Thus the COUP super-flare energy slope is consistent with the previous analyses of COUP flares \citet{Wolk05, Caramazza07, Colombo07, Stelzer07} and the distribution of flare energies found here for MYStIX and SFiNCs PMS stars (\S\ref{sec:energy_distributions}). 

%[ht]

\section{Poisson regression model with multiple changepoints} \label{sec:change_point_model}

We adopt here a statistical model of PMS X-ray variability as a sequence of stepwise constant flux values in a Poisson counting stochastic process.  The statistical problem of multiple changepoint detection in a time series has a substantial history dating back to \citet{Quandt58} with Bayesian approaches starting with \citet{Barry93}. It was an application in the famous paper by \citet{Green95} introducing reversible jump Markov chain Monte Carlo methods.  

A unique optimal solution to the maximum likelihood or Bayesian inference problem exists but, unless the number of partitions is known in advance, it is difficult to obtain as there is a vast number of possible partitions of the time series.  Standard computational procedures (such as the EM Algorithm) cannot be directly applied because the Poisson multiple changepoint model is discontinuous and nonlinear, and models with different numbers of changepoints are not nested. 

This statistical model is widely used in high energy astrophysics under the rubric `Bayesian Blocks' proposed by \citet{Scargle98}. Scargle's original algorithm was approximative and often arrived at significantly suboptimal partitions.  An improved computational methodology based on dynamic programming was developed by Scargle in collaboration with a team of computer scientists \citet{Jackson05} and \citet{Scargle13}.

The computational procedure we use here is based the foundational work of econometrician \citet{Chib98} with developments by statisticians \citet{Fruhwirth06}.  Chib's procedure reparameterizes the changepoint model as a finite mixture model with latent state variables with an unknown number of hidden time delimited regimes.  It is thus an example of `state space modeling', a powerful approach for advanced time series modeling using hierarchical models \citep{Durbin2012}. This formulation is described in less technical language by \citet{Park10}; see also the lectures by \citet{Brandt10}. Applications in the social sciences are presented by \citet{Brandt09} and \citet{Park11}.  

The statistical model for the Poisson regression model with multiple changepoints is:

\begin{equation}
p({\bf \beta, P, s | y)}  \propto \prod_{t=1}^T p(y_t | {\bf \beta, P, s)} ~ \prod_{i=1}^M p({\bf \beta_i}) p({\bf p_i})  \nonumber
\end{equation}
\begin{equation}
 = {\it Poisson} (y_i | {\bf \beta_i}) 
 \times \prod_{t=2}^T \sum_{m=1}^M Poisson(y_t | {\bf \beta_m}) Pr(s_t = m | {\bf \beta, P})  \\
 \times \prod_{i=1}^M Normal({\bf \beta_{0,i}, B_{0,i}}) Beta(a_i,b_i) 
\end{equation}

\noindent Here $y_t$ are the Poisson-distributed counts in time bin centered at time $t$, $m$ is the hidden state at time $t$, $s_t$ are the latent state variables with values $1, 2, ..., M$,  and $\beta_m$ are the regression parameters (flux levels) for each of the $M$ states. The prior choices are a  multivariate normal for the regression parameters and a Beta distribution for the transition probabilities. 

\begin{figure}[ht]
\plotone{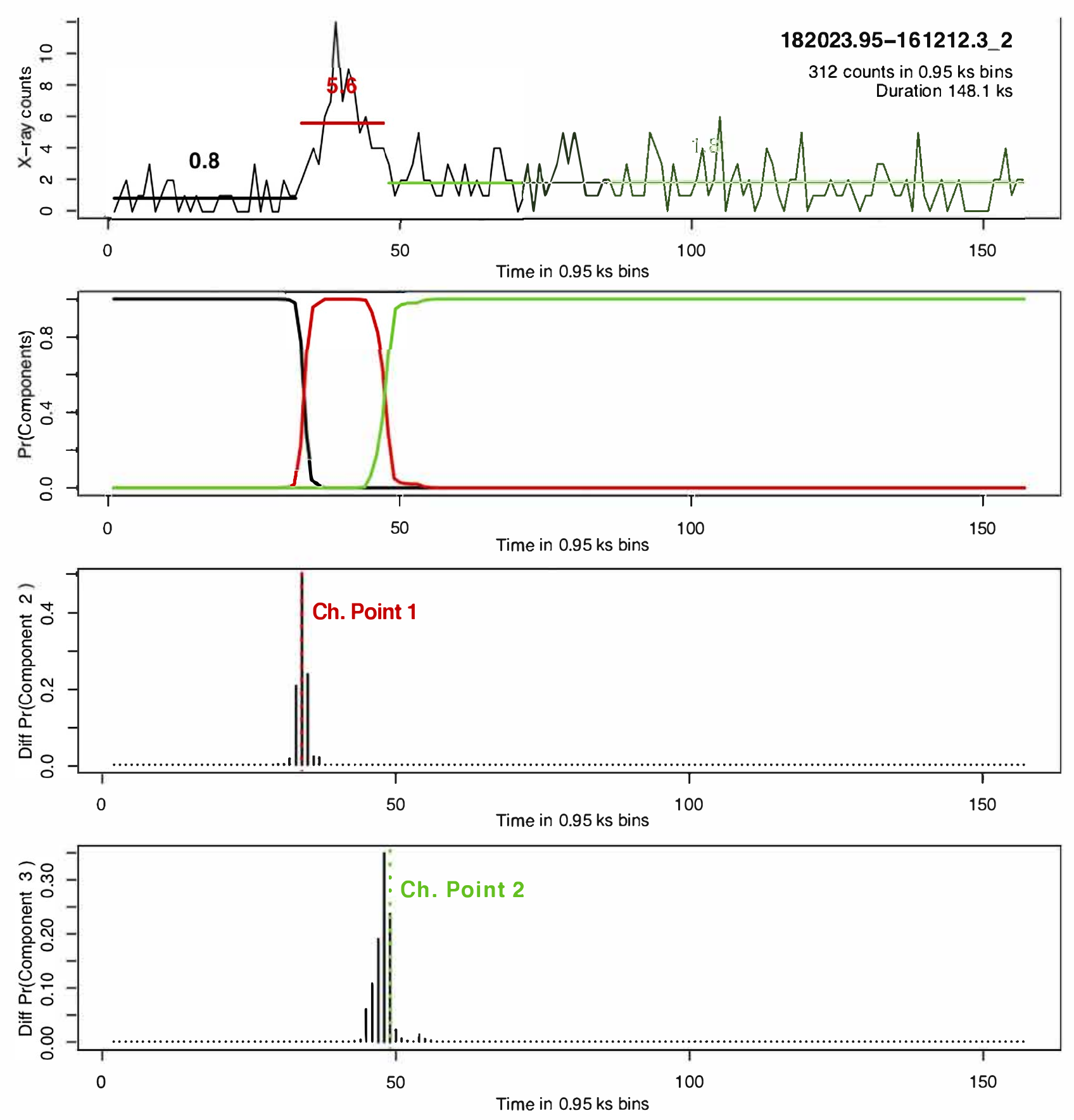}
\caption{Poisson regression model with two changepoints for the second ObsID of {\it Chandra} source 182023.95-161212.3, a pre-main sequence star in the M 17 star forming region. See text for details. Further examples of the top panel are shown in Figure~\ref{fig:flare_classes}. \label{fig:change_points}}
\end{figure}

The calculation proceeds in three steps: hidden state variables are sampled using Chib's recursive algorithm;  transition probabilities to a new state are drawn from binomial distributions; and state space variables and transition probabilities are sampled using a specialized data augmentation algorithm designed for count data.   Software implementation is available in the {\it MCMCpoissonChange} function within the {\it MCMCpack} CRAN package \citep{Martin11} of the R public domain software environment \citep{RCoreTeam20}.  Calculations are performed in R and C++.  Our code and graphics are based on R scripts by \citet{Brandt10}.  

For this model, the X-ray counts must be placed into evenly spaced time bins; this is the data structure required for most methods of time series analysis.  We find that the calculated changepoints are insensitive to the choice of bin width; bin sequences can be dominated by zeros and ones, or can have many counts.  We chose bin widths individually for each source so that bins typically have $\sim 2$ counts for the faintest sources ranging to $\sim 16$ counts for the brightest sources.  Following the default in  {\it MCMCpoissonChange}, we adopt the simple $Beta(1,1)$, or  uniform, prior distribution for the transition probabilities. Results seem resistant to reasonable variations in the prior, but sensitivity to weak flares is reduced for narrow priors like Beta(10,1). We further restrict model complexity to 1 through 5 changepoints, producing 2 to 6 stepwise segments. The model with the highest Bayes factor is chosen.  The chosen model may not be unique; models with fewer or more changepoints may also give satisfactory fits using the Bayes factor significance thresholds recommended by \citet{Kass95}. The algorithm does not permit comparison to a constant (zero changepoint) model.   For weak sources, a spurious short segment is sometimes introduced around the initial bins as the algorithm forces the first point to be zero. 

Figure~\ref{fig:change_points} gives an example of the graphical output for the 3,143 ObsIDs identified in \S \ref{sec:flare_identification}. The top panel shows the binned lightcurve with 312 counts extracted over a 148.1~ks exposure that were placed into bins with width 0.95~ks. Notice that this exposure time is the arrival time difference between the first and last X-ray counts and thus can be slightly smaller than the actual {\it Chandra} observation exposure time (Table \ref{tab:super-flares}), which is the difference between the start and stop times of the observation.  Most bins have $1-3$ counts but a flare is seen midway through the observation.  The best changepoint model has three segments with averages of 0.8, 5.6 and 1.8 counts~bin$^{-1}$.  The second panel shows the relative probabilities of the three segments based on Bayes factors for each bin of the lightcurve.  The third and following panels show the Bayesian posterior probability distribution functions for the segments in the best fit model, omitting the first segment. These are essentially the differentials of the second panel distributions.  

In most cases, we choose the changepoint time to be the bin where the new component has probabilities exceeding 0.5; that is, when it dominates over all other components.  These changepoint times are shown with vertical dashed lines in Figure~\ref{fig:change_points}.  However, visual examination of the lightcurves sometimes shows low-level variations at the beginning of the flare are missed with this criterion.  In these cases, the start time (and more occasionally, stop time) was manually adjusted to include the full flare.  

The \citet{Chib98} formulation of the Poisson multiple changepoint problem, with the software implementation of \citet{Martin11} and \citet{Brandt10}, should give solutions very similar to the Bayesian Blocks formulation of \citet{Scargle13}.  Its graphical output, illustrated in Figure~\ref{fig:change_points}, allows flexible scientific interpretation.  We have chosen changepoints when a component's probability crosses the 50\% boundary, but another user might choose a 90\% criterion for a more conservative evaluation of change.  There are also situations where several components simultaneously contribute to the model; caution might be warranted in selecting changepoints in such cases.  

\section{Derivation of stellar properties}  \label{sec:stellar_props_appendix}

New parallax-based distances are obtained for many MYStIX and SFiNCs star forming regions by  \citet{Cantat-Gaudin18,Kuhn19} and other recent {\it Gaia}-based studies. Details are given in Table~\ref{tab:mystix_sfincs_regions}. These new distances allow us to re-calculate X-ray luminosities from {\it Chandra} fluxes, which are derived from {\it Chandra} count rates and median energies using the scalings of \citet{Getman10}.  

Preliminary PMS stellar masses are estimated from the empirical scaling relation between X-ray luminosity and stellar mass obtained for the nearby Taurus star forming region with the {\it XMM} X-ray telescope \citep{Telleschi07}.  

For solar and lower mass stars, visual band absorptions, $A_V$, are estimated by dereddening near-infrared colors to the intrinsic color locus of the Taurus low-mass stars in the $J - H$ versus $H - K_s$ color–color diagram.  Details of the procedure appear in \citet{Getman14a}. 
%%For higher mass stars, $A_V$ is estimated by averaging the source extinctions of a dozen PMS stars closest on the sky.
\begin{figure}[ht]
\plotone{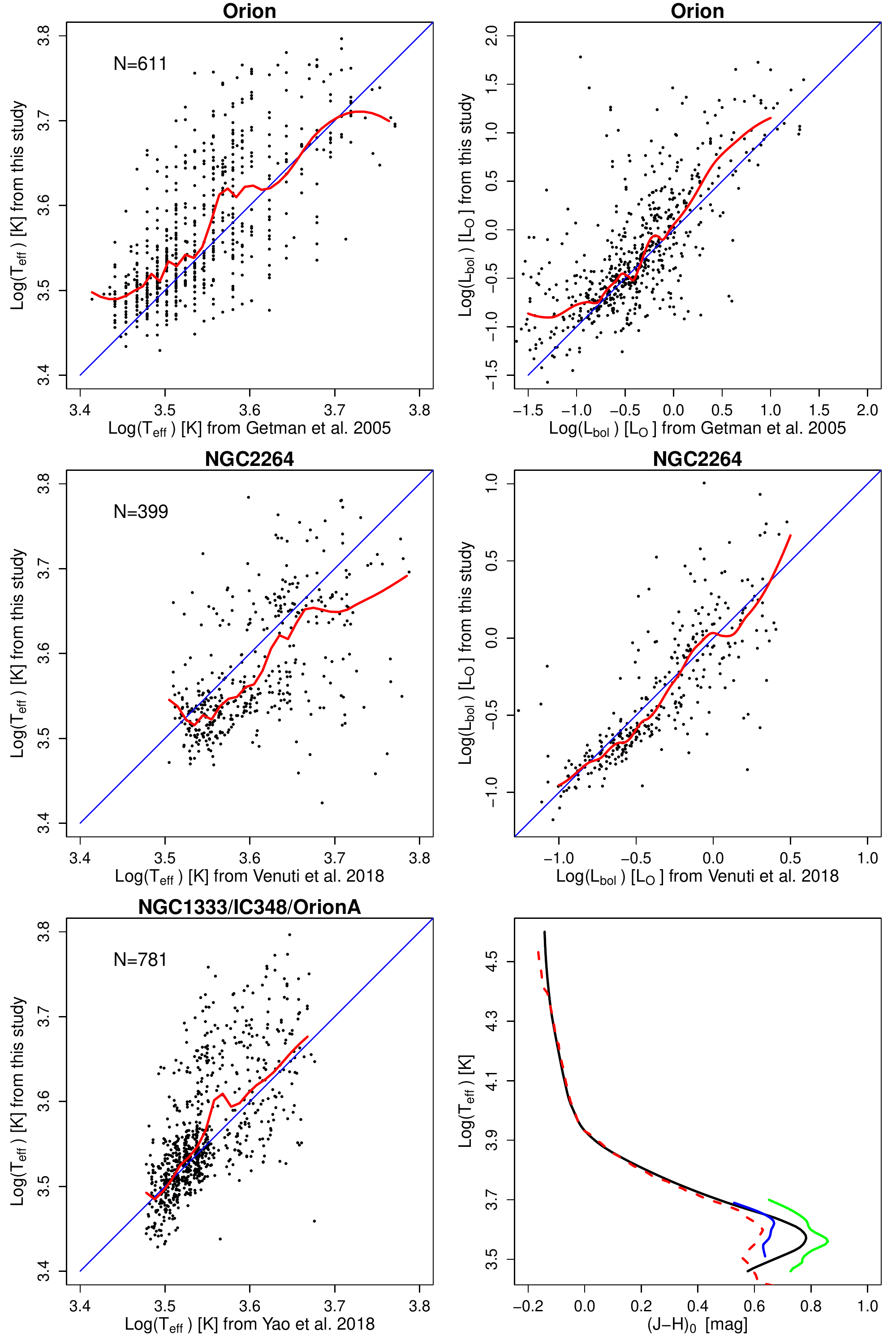}
\caption{Comparison of $T_{eff}$ and $L_{bol}$ estimates for individual stars in selected star formation regions. The ordinate gives values derived in this study based on X-ray and infrared photometry, while the abscissa gives values from a published study based on optical spectroscopy (see text for details). Blue lines show equal values, and red curves show local regression fits to the median ordinate values. The bottom-right panel presents the temperature-color scales for young stars in four studies; see text for details. The legends give the numbers of involved stars.  \label{fig:teff_spectroscopy_vs_photometry}}
\end{figure}

Ages of these low-mass MYStIX/SFiNCs young stars are estimated in two ways.  For stars with $\geq 10$ X-ray source photons, the $Age_{JX}$ X-ray/near-infrared chronometer developed by \citet{Getman14a} is used. In this method, the X-ray luminosities (as surrogates for stellar masses using the Telleschi et al. relation) and $J$-band magnitudes corrected for extinction (as surrogates for bolometric luminosities) are combined with PMS evolutionary models to obtain individual stellar ages. Inferred $Age_{JX}$ for MYStIX/SFiNCs stellar clusters are consistent with independently known patterns of star formation histories, cluster sizes, and cluster disk fractions \citep{Getman14a,Kuhn15b,Richert18}. In our recent work by \citet{Richert18} such ages were calculated for the three different evolutionary models with different treatments of interior magnetic fields  \citep{Siess00,Choi16,Feiden16}. 
In the current work, we choose to recompute ages using the popular PARSEC 1.2S evolutionary models \citep{Bressan12,Chen14}. The empirical changes to the relationship of the temperature and mean optical depth across a stellar atmosphere implemented in these models may mitigate the ``radius inflation'' problem related to effects of surface starspots and interior magnetic pressure on the stellar structure, slowing convective energy flux and global contraction, and changing the location of PMS isochrones in the Hertzsprung-Russell (HR) diagram \citep{Morrell19}.

For low-mass stars without $Age_{JX}$ estimates (due to very weak X-ray sources or inaccurate near-infrared photometry), and all intermediate- and high-mass stars, we assign age values to be the median ages among seven nearest (on the sky) low-mass $Age_{JX}$ neighbors. For cases when young stellar objects are members of highly embedded MYStIX/SFiNCs subclusters without known $Age_{JX}$ stellar members, age estimates are obtained from the $Age_{JX} - (J-H)$ relationship \citep{Getman14a} transformed to the PARSEC 1.2S scale.  These ages are truncated at $\ga 0.4$~Myr (due to the paucity of $<0.4$~Myr sources on the age-color diagram) and  $\la 5$~Myr (due to the degeneracy of PMS isochrones on the $M_{J} - L_{X}$ diagram).  

On the $J$ versus $J-H$ diagram (which is insensitive to the presence of circumstellar disks), we de-redden each MYStIX/SFiNCs star with reliable photometry towards the intrinsic color-magnitude locus corresponding to a PARSEC 1.2S isochrone of the star's estimated age.  This gives estimates of $A_V$, stellar $T_{eff}$, $L_{bol}$, $R$, and $M$. Such estimates are obtained for $>$26,000 out of $\sim$40,000 MYStIX+SFiNCs young stellar objects.  

Due to the degeneracy of PMS isochrones at the intermediate-mass range, the inferred properties of some stars are associated with ranges of values rather than single values. To overcome this problem, a few additional steps were taken. First, all bright ($J<15$~mag) MYStIX+SFiNCs stars are passed through the Virtual Observatory Spectral energy distribution Analyzer \citep[VOSA;][]{Bayo2008} using (in addition to our $JHK_s$ and $Spitzer$-IRAC photometry) data from numerous other optical and IR photometric catalogs, such as {\it Gaia}-DR2 \citep{Gaia2016,Gaia2018}, Pan-STARRS \citep{Chambers2016}, SDSS \citep{Alam2015}, VPHAS-DR2 \citep{Drew2016}, APASS-DR9 \citep{Henden2015}, CMC14 \citep{CMC2006}, Tycho-2 \citep{Hog2000}, VVV-DR2 \citep{Minniti2017}, DECam \citep{DePoy2008}, and others.  These star's SEDs were fit with the BT-Settl atmospheric model \citep{Allard2012} providing independent $T_{eff}$ and $L_{bol}$ estimates. Second, the $JHK_s$- and VOSA-based outcomes are compared with each other. Third, for stars in several MYStIX+SFiNCs regions with published optical-IR spectroscopy \citep{Getman05, Skiff2014,Venuti2018, Yao2018},  their effective temperature and bolometric luminosity are compared with the photometric outcomes of both the $JHK_s$- and VOSA-based methods. These extra steps allow selection of most likely $T_{eff}$ and $L_{bol}$ solutions among the $JHK_s$ and $VOSA$ choices for intermediate-mass stellar candidates. Most of these solutions are based on the VOSA modeling.

Specifically, X-ray/NIR-derived and VOSA-derived $T_{eff}$ and $L_{bol}$ quantities are available for 26,681 and 12,183 (with $J<15$~mag) MYStIX/SFiNCs stars, respectively. In the range $\log(T_{eff}) = [3.4-3.8]$~K (roughly $[0.1-3]$~M$_{\odot}$), the median and Inter Quartile Ranges (IQRs) of the $\log(T_{eff;xrayNIR}) - \log(T_{eff;VOSA})$ differences are 0.02 and 0.08, respectively. This $\log(T_{eff})$ range corresponds to the $\log(L_{bol})$ range of $[-1.5$~to~$2]$~L$_{\odot}$. The median and IQR of the $\log(L_{bol;xrayNIR}) - \log(L_{bol;VOSA})$ differences are -0.04 and 0.32, respectively. Hence, the distributions of the $T_{eff}$ differences typically have small biases ($5$\%) and dispersions ($10$\%). The distributions of the $L_{bol}$ differences have small biases ($10$\%) but high spreads ($100$\%).

For the regions with published optical-IR spectroscopy, in the intermediate-mass range ($\log(T_{eff} = [3.8-4.0])$~K), where the PMS isochrones on the $J$  versus $J - H$ diagram are degenerate, the VOSA-derived $T_{eff}$ and $L_{bol}$  quantities are consistent reasonably well with those inferred from the optical-IR spectroscopy. For 130 MYStIX/SFiNCs X-ray stars, which lie in this degeneracy locus and have highly uncertain properties obtained with the X-ray-NIR method, their final, chosen stellar properties are those derived with the VOSA method. There are also 23 X-ray stars that lie in this degeneracy locus but have unique solutions inferred using the X-ray-NIR method itself. Overall, among the 26,681 MYStIX/SFiNCs stars with available $T_{eff}$ and $L_{bol}$ estimates, 130 and 26,551 stars have their final (used in \S \ref{sec:flare_vs_mass}) properties obtained with the VOSA and X-ray-NIR methods, respectively.

Figure \ref{fig:teff_spectroscopy_vs_photometry} compares the stellar effective temperatures and bolometric luminosities emerging from this analysis with previously published $T_{eff}$ and $L_{bol}$ values based on optical spectroscopy for several nearby MYStIX+SFiNCs regions.  Orion Nebula Cluster values are compared to \citet{Getman05}, NGC~2264 values are compared to \citet{Venuti2018}, and NGC~1333, IC~348, and Orion~A values are compared to \citet{Yao2018}. The comparison of red and blue curves shows that the bias is typically  $<200$~K in $T_{eff}$ and $0.1$~dex in $\log(L_{bol})$. However the scatter of individual stars is larger with IQRs around $\pm 500$~K in $T_{eff}$ and $\pm 0.4$~dex in $\log(L_{bol})$. 

The bottom-right panel of Figure \ref{fig:teff_spectroscopy_vs_photometry} compares the color-$T_{eff}$ relation for four studies. The black curve is from the PARSEC 1.2S model used in this study.  The comparison green curve is from \citet{Getman05} (green), the blue curve is from \citet{Venuti2018}, and the red curve is from the popular in the recent literature transformation by \citet{Pecaut2013}. This diagram shows that there is little disagreement between several color-temperature transformations for solar- and intermediate-mass stars, but discrepancies are present for cooler stars. Such discrepancies may contribute to the biases seen between our and spectroscopic-based estimates for Orion and NGC 2264 stars.

We conclude that the stellar properties derived here, based on X-ray and $JHK_s$ photometry and modern stellar interiors models that account (directly or indirectly) for the effects of magnetic fields, are reasonably accurate to obtain trends in stellar properties related to X-ray super-flare occurrence (\S\S \ref{sec:allmystixsfincs_vs_super-flares} and \ref{sec:flare_vs_mass}).  Considerable spreads are present in the $T_{eff}$ and $L_{bol}$ values, but we recall that $\sim 500$~K systematic uncertainties for $T_{eff}$ values in PMS stars can be present even in spectroscopic studies \citep[see Figure~3a in][]{Yao2018}.  

Following \citet{Richert18}, apparent (non-dereddened) {\it Spitzer}-IRAC spectral energy distribution (SED) slopes, $\alpha_{IRAC} = d \log(\lambda F_{\lambda})/d \log(\lambda)$, measured in the IRAC wavelength range from 3.6 to 8.0~$\mu$m are employed to distinguish between disk-bearing ($\alpha_{IRAC} \leq -1.9$) and diskless ($\alpha_{IRAC} > -1.9$) stars. 

All these inferred properties of the X-ray super-flare host stars are given in Table~\ref{tab:flare_host_properties}. 

%%%%\newpage
\bibliography{ms_bibliography}{}
\bibliographystyle{aasjournal}
\end{document}